# A novel brain partition highlights the modular skeleton shared by structure and function

Ibai Diez[1,*], Paolo Bonifazi[2,*], Iñaki Escudero[1,3], Beatriz Mateos[1,3], Miguel A. Muñoz[4], Sebastiano Stramaglia[1,5,6,**,§] and Jesus M Cortes[1,6,7,§]

1. Biocruces Health Research Institute, Cruces University Hospital, Barakaldo, Spain.

2. School of Physics and Astronomy, George S. Wise Faculty of Life Sciences, Sagol School of Neuroscience, Tel Aviv University, Tel Aviv, Israel.

3. Radiology Service, Cruces University Hospital, Barakaldo, Spain.

4. Departamento de Electromagnetismo y Física de la Materia and Instituto Carlos I de Física Teórica y Computacional, Universidad de Granada, Spain.

5. Dipartimento di Fisica, Universita degli Studi di Bari and INFN, Bari, Italy.

6. Ikerbasque: The Basque Foundation for Science, Bilbao, Spain.

7. Department of Cell Biology and Histology. University of the Basque Country. Leioa, Spain

* These authors contributed equally to this work.

** New address: BCAM - Basque Center for Applied Mathematics, Bilbao, Spain

§ To whom correspondence should be addressed.

Email: Sebastiano.Stramaglia@ba.infn.it and jesus.cortesdiaz@osakidetza.net

**Abstract**

Elucidating the intricate relationship between brain structure and function, both in healthy and pathological conditions, is a key challenge for modern neuroscience. Recent progress in neuroimaging has helped advance our understanding of this important issue, with diffusion images providing information about structural connectivity (SC) and functional magnetic resonance imaging shedding light on resting state functional connectivity (rsFC). Here, we adopt a systems approach, relying on modular hierarchical clustering, to study together SC and rsFC datasets gathered independently from healthy human subjects. Our novel approach allows us to find a common skeleton shared by structure and function from which a new, optimal, brain partition can be extracted. We describe the emerging common structure-function modules (SFMs) in detail and compare them with commonly employed anatomical or functional parcellations. Our results underline the strong correspondence between brain structure and resting-state dynamics as well as the emerging coherent organization of the human brain.

Two complementary principles underlie human brain functioning, segregation and integration (1). Segregation refers to the need of functionally specialized brain areas to process incoming information and to perform distinct tasks separately, whereas integration or binding is required for the coordinated activation of neuronal populations across brain areas that result in coherent cognitive and behavioral states (2,3). Elucidating the wiring architecture of brain networks is essential to understanding how an optimal balance between segregation and integration might be achieved, and it constitutes a key challenge in contemporary neuroscience.

At least three different classes of brain networks can be studied[1] (4-5): "structural connectivity" (SC) networks, encoding anatomical connections or links between neural elements or brain regions (6); "functional connectivity" (FC) networks, defining the activation profiles among distinct neuronal populations (7); and "effective connectivity" (EC) networks, identifying causal interactions underlying temporally ordered activation or information flow (8).

The development of novel neuroimaging techniques and in particular, advances in magnetic resonance imaging (MRI), have enabled functional brain networks to be monitored and reconstructed, for example, as inferred from correlations between blood oxygen-level dependent time-series (9). Likewise, structural networks have been obtained from diffusion tensor images and high-resolution tractography[2] (10).

---

[1] The three classes of brain connectivity can be described at several levels: at the microscale the associations refer to synaptic connectivity; the mesoscale corresponds to a spatial resolution of hundreds of micrometers; and the macroscale represents a very large number of neuronal populations forming distinct brain regions that are interconnected by inter-regional pathways. Further details in (4).

[2] The combined use of both anatomical connectivity and dynamic models of neural activity to relate both SC and FC has grown over recent years (11-16).

Empirical progress has been complemented with the development of theoretical and computational advances using graph theory (17-19). In particular, complex network analysis has emerged out as a successful framework to scrutinize brain architecture and the relevant features of its emergent dynamic or functional states (20). These studies have rapidly revealed that both SC and FC networks exhibit a hierarchical organization of distinct brain modules (or areas) that communicate through connector hubs (21-23). Hence, it was proposed that such a hierarchically compartmentalized organization is essential for segregation, while the existence of connector hubs and the presence of weak links between otherwise separated moduli facilitates the optimal balance between integration and segregation (22,24).

From a more general perspective, it seems clear that neural dynamics and any associated functions are necessarily constrained by the underlying wiring structure (25), although the precise relationship between SC and FC networks is still far from clear. One key problem when attempting to define such a relationship is that structure-function matching is actually a one-to-many mapping, meaning that for a given fixed anatomy, the functional repertoire needs to be vast in order to facilitate perception, action, memory, cognition and complex behaviors. Thus, bridging the gap between structure and function, and understanding how such a huge repertoire of functional brain states can emerge out from a fixed structure is one of the fundamental challenges in neuroscience (see e.g. (26) and references therein).

As a first step in this direction, several works have analyzed functional connectivity in the resting state, i.e. when the brain being monitored is not involved in any goal-oriented tasks and thus, it is as close as possible to doing nothing. These analyses revealed consistent and robust "resting state" functional connectivity (rsFC) patterns across subjects (27-29). Functional pairwise correlations turn out to be relatively strong between structurally connected nodes (30-36), yet more surprisingly, strong functional connections also appear commonly between distant regions that lack direct structural connections[3] (32,34), revealing the existence of strong "indirect effects". This observation suggests that over and beyond direct node-to-node and link-to-link pairwise comparisons, more collective or systemic analyses will be necessary to shed light on the relationships between SC and rsFC networks.

We propose here, following recent work adopting the same strategy (37-39), to shift attention to groups of nodes, and to contrast structural and functional networks by exploiting their hierarchical modular organization. More specifically, by employing the template of hierarchical modular organization derived from structural data to represent the resting state functional one and vice versa, we search for the optimal common partition shared by structure and function by maximizing a novel quantity, that we dub "cross-modularity". Our hypothesis is that, if we assume that segregated functions are associated with distinct structural moduli, visualizing the rsFC data in terms of the natural structural moduli derived solely from network architecture (i.e. the SC) should help define and highlight how strongly structure constrains function. Conversely, the functional hierarchical-modular organization can be employed to visualize structural data. As such, these two complementary approaches should shed light on the intricate relationships between structure and function, and in particular, this procedure allows the extraction of an optimal partition illustrating that structure and function are much more tightly correlated than previously thought. The novel partition that we uncover here divides the brain into disjointed regions that we refer to as common "structure-function modules" (SFMs), representing a coarse-grained skeleton of the brain, which is largely shared by structure and function.

**Results**

---

[3] To overcome this limitation, the authors in (34) first made a suitable selection of brain regions to calculate the correlation on individual links in those masks, showing an enhancement of SC-rsFC correlations.

**SC and rsFC networks**

We obtained SC and rsFC networks from same-subject DTI and rs-fMRI data gathered from a population of healthy human subjects (n=12, age 33.5 ± 8.7 years old). The data was obtained at a resolution of 2514 regions of interest (ROI), enabling SC and rsFC (2514*2514) matrices to be derived for each subject (see methods). All the 2514 ROIs have a variable size (figure S9, solid line); different sizes range from values between 100 mm3 to 3400 mm3 (mean=535.7, SD= 237.4). Even more variable are the ROI sizes within cerebellum (mean=909.7, SD=560.5) (figure S9, dashed line).

Although the two datasets are acquired from the same subject, it is noteworthy that the two (rsFC and SC) networks are obtained independently, and they constitute two separate and autonomous datasets correlated to different physical and physiological mechanisms, and corresponding to distinct measures acquired and post-processed in a different manner. To focus on generic aspects and not on individual singularities, we obtained the mean SC and rsFC networks by averaging the individual matrices of the subjects using adequate templates and projections, as described in methods.

**Link-to-link comparison**

As a preliminary analysis, following standard approaches, we measured the Pearson correlation coefficient (r) between the averaged structural and functional matrices (31,33). This provided a link-to-link comparison between the two networks and gave an overall value of about r=0.2 (see figure S1), indicative of a rather weak correlation. By restricting the measurement to existing physical links, this value increased to a moderate value of r=0.3 and it can be augmented a little further by comparing only pairs of nodes within (but not between) structural moduli (see below for a proper definition and evaluation of moduli). In any case, the Pearson coefficient value remained below r=0.5, even having discarded a large fraction of (between-module) links. We also measured the "similarity" (L) between the two matrices, a measure of the fraction of elements that are simultaneously above or below variable thresholds in both of them (see methods). We obtained a moderate value of L=0.45 (figure S2a), which can be slightly increased at the cost of restricting the comparison to within-module matrix elements. In conclusion, this type of element-to-element matrix comparison reflects moderate levels of similitude between SC and rsFC networks.

**Comparison at module level**

In order to extend the comparisons between structural and functional brain networks, we analyzed the two networks at the moduli level (figure 1a), applying a standard hierarchical agglomerative clustering (HAC) algorithm to the mean rsFC network (40-42). This approach enables a hierarchical tree or dendrogram (figure 1c) to be constructed in which nodes are progressively merged together into moduli following a nested hierarchy of "vicinity" (which reflects correlation in the case of rsFC data). Cutting this tree at any arbitrary level leads to a pooling of the initial 2514 ROIs into a finite number of modules (M) that can be tuned by varying the depth of the cut. For instance, the partition with maximal Newman's modularity Q, (i.e.: with a maximal fraction of intra-module to inter-module links with respect to randomizations) of the rsFC matrix corresponds to division into about 12 disjointed communities, although similar modularity Q values were obtained for divisions of M from 12 up to 25 (figure S2b).

Next, to scrutinize the averaged SC matrix we considered each of these functional partitions, with progressively larger values of M, separately. For instance, the partition into 20 moduli (obtained from the srFC data and illustrated in figure 1) provides a remarkably large value of modularity Q for the SC matrix. In other words, a partition inferred exclusively from functional measurements leads to an excellent ordering of the structural/wiring data, allowing for a good organization and visualization of structural moduli (figures 1a and 1b). Conversely, employing an optimal partition for the SC network (i.e., the one with maximal Newman's modularity Q) entails quite a large modularity for the averaged rsFC matrix (figure S2b). This simple observation constitutes an important finding: both rsFC and SC networks display high modularity (as already acknowledged) but with a previously unnoticed yet excellent match between functionally and structurally identified modules. As an example to illustrate the excellent matching between structure and function, figure 1e and movie S22 show a single module; it consists of different functionally correlated subregions (marked in red) which apparently are physically far from each other; however, they can be observed to be wired together by fiber bundles, forming a coherent, though de-localized module.. Similarly, all the M=20 modules are represented in figures S3 and S4. This observation, structural modules matching functional ones, implies that most of the aforementioned indirect effects observed in the resting-state reflect functional correlations that stem from the existence of modules[4]. That is, most of the functional pairwise node-to-node correlations that cannot be explained by direct structural connectivity can be accounted for by their corresponding nodes lying in structurally connected moduli, even if the specific nodes lack direct wiring.

**Cross-modularity index and SFMs**

To quantify the striking observation that a common partition into modules or communities describes both the rsFC and SC data remarkably well (figure 1), we introduced an index $X$ called "cross-modularity" (see methods), which is large for a given partition if the corresponding Newman's modularities of the two matrices under comparison are large and there is also a large within-module similarity between both divisions (i.e., a large fraction of existing intra-module links are shared by both networks). Thus, a large cross-modularity value indicates that, using a given common partition, both matrices are highly modular and, at the same time, the moduli are internally wired in a similar way. A maximization of $X$ across possible partitions allows the finding of an optimal structure-function brain partition. Indeed, through this novel index, we found that the partition into 20 moduli derived from rsFC data (as portrayed in figure 1) is optimal (figure 2), although similar quality partitions can be obtained in the range of M from 10 to 30. The reason why the cross-modularity index is almost constant in this interval is that there is an overall balance between two opposing effects, the both of them occurring when M increases: (i) the increase in the similarity between SC and rsFC (see figure S2a) and (ii) the decrease in Newman's modularity produced by an increase of inter-module connections and a decrease of intra-module connections (figure S2b).

---

[4] To illustrate this effect, one might consider two completely isolated moduli each of them representing a fully connected network with a large number of nodes, and imagine adding a single strong link between two given nodes, one from each module. Even if there is just a single structural connection, it is very likely that any dynamics running on top of this structure will generate node-to-node correlations between non-directly connected inter-module pairs of nodes. However, this functional correlation would not be detected by a straightforward pairwise comparison using the structural matrix but rather, it emerges naturally once an ordering is performed in terms of dominant moduli.

The modular decomposition of the brain proposed here is quite robust across subjects: evaluating the corresponding cross-modularity, using each individual SC, leads to similar patterns to those obtained using the SC averaged over the the N=12 subjects. Moreover, it was also notable that slightly larger cross-modularity values (about 4% higher) were obtained using moduli derived from the rsFC to study the SC than vice versa, thus, we focus on the first choice.

The brain partition for M=20 is described anatomically in Table S1, illustrated in figures 3 and 4 and movies S1-S20. When looking at the spatial distribution of SFMs (figures 3 and 4), a high degree of symmetry exists between the two hemispheres in most of the modules (e.g., modules 3, 6 and 12: see also movie S21 for a 3D superposition of the 20 modules). Observe, for example, in figures S3 and S4, that moduli composed of segregated islands –characterized by correlated functional activity-- have always structural connections bridging them, thus providing the possibility of functional cohesion.

Remarkably, the agreement between rsFC and SC was systematically better when the averaged across-subject matrices were used rather than those obtained on single subjects, which nevertheless remained high (figure

S5). This illustrates the robustness of the obtained partition, which is preserved despite of the existence of individual specific traits.

## Discussion

### Anticorrelations in SFMs

Whilst all the moduli in SFMs appear to be internally correlated, strong inter-module anti-correlations also exist with a pronounced modular structure (figure 1a). This observation becomes even more evident when plotting just the sign of the functional correlation in SFMs (figure 5): red for positive correlations; blue for anti-correlations; and green for values close to zero (i.e.: ranging between -0.1 and 0.1). This not only reveals that all SFM resting-state moduli are internally correlated but many of them tend to be anti-correlated with others. In particular, our results showed that modules 9 and 10 are positively correlated with one another, and that both modules are strongly anti-correlated with module 3 (and more weakly with other modules), a module that strongly overlaps with the sensory-motor task-related network (see figure 6b). Whilst the existence of anti-correlations in resting state networks (over and above processing artifacts) has been subject of some debate (33), it is now well-established that anti-correlations are inherent to resting state functional networks. In particular, it has been shown that the strongest anti-correlations are mediated by the default mode network (DMN), particularly with task performing areas (43). This is indeed consistent with our results, as modules 9 and 10 (those with the strongest anti-correlations) overlap significantly with the DMN (see figure 6b and Table S1).

### Overlap between SFMs and AAL brain partition

We asked whether SFMs bore any resemblance (measured as percentage overlap, see methods) with other macro-scale brain parcellations commonly found in the literature, which typically are based solely either on function or on structure (whilst SFMs aim at describing both). First, we analyzed the overlap between SFMs and the brain regions belonging to a structural atlas, the automated anatomical labeling (AAL) (44), assessing the results for the 45 AAL homologue areas[5] and for our 20 modules partition (i.e., SFMs, figures 6a). Examining the overlap between SFMs and AAL brain areas, we found that different anatomical areas are included in a single SFM (i.e., a column of the matrix), clearly highlighting how SFMs (that underlie different brain states at resting conditions) simultaneously recruit distinct brain circuits. Conversely, the same area of the AAL (the row of the matrix in figure 6a) might be included in several SFMs, highlighting the anatomical overlap of the latter. Such observations do not depend significantly on the number of modules composing the partition, since the overlap between a single AAL area and a single SFM never approaches a unitary value (figure S6a). Some AAL areas evidently have a strong overlap with the M=20 modules (e.g., SFM 11 and the Rectus gyrus, or SFM 16 and the Temporal middle gyrus, figure 6a), indicating that such anatomical areas might have a much more relevant functional role than previously believed in comparison to other AAL areas. Moreover, both the Rectus gyrus and Middle temporal gyrus seem to be functionally represented on a smaller scale as the overlap between them remain high as the number of modules increases (figure S6a). Other areas are also functionally represented on a smaller scale (e.g., the Thalamus, approx. M=40 onwards, figure S6a).

**Overlap between SFMs and RSNs**

Similarly, we compared the well-studied Resting State Networks[6] (RSNs), a brain functional atlas constructed using independent component analysis of functional data (27-29,45), to the M=20 brain partition of SFMs

---

[5] The AAL (44) is a well-known anatomical atlas in which, the brain (after removal of the cerebellum) is divided into 90 ROIs, including cortical and sub-cortical regions like the hippocampus, amygdala and thalamus. Each anatomical region is localized within the two hemispheres, such that the hippocampus is divided in two ROIs, one in the left hemisphere and the other one in the right, as is the thalamus and so forth. Thus, the 90 ROIs can be grouped in 45 homologues areas, folding the left and the right ROI from each hemisphere.

[6] RSNs arise from the correlation in signal fluctuations across brain regions and they are a pivotal element in understanding the dynamics and organization of basal brain activity, both in health and disease (27-29). RSNs are observed during the resting state, a condition defined by the absence of goal-directed behavior or

(figure 6b). Two RSNs display a strong overlap (>0.5) with two distinct SFMs, the "Sensory motor" with SFM 3 and "Medial visual" with SFM 4, while the other RSNs overlap with more than one SFM. Irrespective of the number of modules imposed on the FC matrix (figure S6b), we did not observe a complete overlap between RSNs and SFMs, in part due to the fact that SFMs are distinct to RSNs but also, to the large inter-subject variability that exists in the shape of each individual RSN. Similar results to those for the AAL and RSNs were also observed but for Brodmann areas, characterized by known neuro-psychological functions, figures S7. Indeed, Brodmann area number 18 is well-characterized by SFM 4 and Brodmann area 20 matches to SFM 18 (figure S7a). Moreover, when the number of the modules in the partition increases, Broadmann areas 18, 20, 11, 10, 19 and 21 are represented by SFMs (figure S7b).

Therefore, we conclude that SFMs: (i) represent a distinct brain partition from those previously described in the literature; (ii) incorporate distinct both structural and functional brain regions into a single operative network/unit; and (iii) can overlap and share both anatomical and functional brain regions (see also Table S1). In the same way as alterations in resting state networks have been reported in several brain pathologies and diseases[7], we expect that the use of the new brain partition represented by SFMs, with simultaneous focus on structure and function, might help also in diagnosing disease.

**SFMs validation with data from the Human Connectome Project**

We want to emphasize that the new brain partition discovered here using our own data recorded in the Cruces University Hospital (Bilbao, Spain) has been fully validated with data by the WU-Minn Human Connectome

---

salient stimuli. Despite the simplicity of the context in which these brain activity patterns are generated, the RSN dynamics are rich and complex. Different RSNs have been associated to specific cognitive representations, e.g., there are visual networks, sensory-motor, auditory, default mode, executive control and some others (for further details see for instance (45).

[7] Previous studies found alterations in RSNs in brain pathological conditions such as in patients with deficit of consciousness after traumatic brain injury (46-49), schizophrenia (50-51) and epilepsy (52).

Project (53), released on Jun 2014. Indeed, maximization of the cross-modularity, on data from the WU-Minn Human Connectome Project, leads to a very similar brain partition and the optimal solution is in concordance with the one reported in this manuscript (figure S8).

**Final considerations**

We hypothesize that since we are looking at the same entity, i.e. the brain, it is reasonable to expect that a brain partition (common to both structure and function) might exist, but to the best of our knowledge, such partition has not yet been found. Indeed, different authors (11-16, 31-33) have shown that brain activity cannot be simply inferred from the underlying structural network of interconnections, i.e. that functional and structural networks are very different objects. On the other hand, recent studies suggest a stronger relationship between the structural and functional network. For instance, both of them have been found to share a strong rich-club structure, meaning that moduli are interconnected through some local hub connectors and that such hubs are highly connected among themselves (54). Another recent work has emphasized that resting-brain functional connectivity can be predicted by analytic measures of network shortest communication pathways (55), which strongly support that SC and rsFC are highly related to one another.

The partition we have elucidated here is the one maximizing cross-modularity; however, there is a band with nearly constant values of the cross-modularity such that all the partitions within it are also plausible. Thus, in that plateau slightly different partitions (with 10 to 30 moduli), can be inferred, all of them describing similarly the structure-function interplay.

In our analysis, the asymmetry between the two strategies, structure following function (SF) and function following structure (FS), comes from the intrinsic differences between the two data sets (e.g., the structural network is sparse, whilst the functional network is dense). A more integrated method, considering both sources of recording together, would deal with the general problem of developing effective algorithms to optimize cross-modularity (e.g., by modifications of existing methods for modularity optimization); although we are aware this is a challenging and interesting problem, it is beyond the aim of the current work.

It is important to emphasize that here, we have applied a data-driven approach, and no further assumptions have been made to obtain the novel brain partition. Two reasons might justify the differences we found in comparison with other existing partitions. First, and more importantly, most of the previous approaches considered the number of subcortical regions either absent or accounting for less than 20% of all ROIs, see (10) and references therein. Here, we have incorporated both rsFC and SC data belonging to all subcortical structures (including amygdala, hippocampus and cerebellum). Second, all previous partitions were obtained looking solely at either SC or rsFC data but we, for the first time, have integrated the two data sets to force modules in the brain partition to be relevant to both structure and function.

In summary, our results show that when trying to correlate brain structure with function, a clear structure-function matching emerges when applying a hierarchical modular approach; that is, pooling brain regions of interest into more densely connected modules rather than scrutinizing the similarity at the level of individual links.

**Material and Methods**

**Same-subject structure-function acquisitions**. This work was approved by the Ethics Committee at the Cruces University Hospital; all the methods were carried out in accordance to approved guidelines. A population of n=12 (6 males) healthy subjects, aged between 24 and 46 (33.5 ± 8.7), provided information consent forms before the magnetic resonance imaging session. For all the participants, we acquired same-subject structure-function data with a Philips Achieva 1.5T Nova scanner. The total scan time for each session was less than 30 minutes and high-resolution anatomical MRI was acquired using a T1-weighted 3D sequence with the following parameters: TR = 7.482 ms, TE = 3.425 ms; parallel imaging (SENSE) acceleration factor=1.5; acquisition matrix size=256x256; FOV=26 cm; slice thickness=1.1 mm; 170 contiguous sections. Diffusion weighted images (DWIs) were acquired using pulsed gradient-spin-echo echo-planar-imaging (PGSE-EPI) under the following parameters: TR = 11070.28 ms, TE = 107.04 ms; 60 slices with thickness of 2 mm; no gap between slices; 128x128 matrix with an FOV of 23x23 cm. Changes in blood-oxygenation-level-dependent (BOLD) T2* signals were measured using an interleaved gradient-echo EPI sequence. The subjects lay quietly for 7.28 minutes, during which 200 whole brain volumes were obtained under the following parameters: TR = 2200 ms, TE = 35 ms; Flip Angle 90; 24 cm field of view; 128x128 pixel matrix; and 3.12 x 3.19 x 4 .00 mm voxel dimensions.

**Data preprocessing.** To analyze the *diffusion weighted images* we first applied the eddy current correction to overcome artifacts produced by changes in the gradient field directions of the MR scanner and subject head movement. Using the corrected data, a local fitting of the diffusion tensor was applied to compute the diffusion tensor model at each voxel. Subsequently, a FACT (fiber assignment by continuous tracking) deterministic tractography algorithm (56) was employed, by using an interactive software for fiber tracking called "Diffusion Toolkit" (57). Tractography algorithms were developed to reconstruct white matter pathways in the brain –connecting grey matter regions-- from diffusion tensor imaging (DTI) data. The FACT algorithm reconstruct individual fibers and tracks them by connecting the voxel where the fiber is initiated with the adjacent one toward which the fiber direction (as determined by the leading local eigenvector of the diffusion tensor), and by iterating this procedure until it is terminated according to the criterion that the fiber arrives to a grey matter region (as identified by a fractional anisotropy index equal to 0.1, characteristic of grey matter). An additional termination criterion is that we avoided sharp curvatures of axonal tracts by fixing a maximum angle variation of 35 degrees from a given voxel to the following one (for further technical details on the employed tractography algorithm see (58).

The *functional MRI* data was preprocessed with FSL (FMRIB Software Library v5.0). The first 10 volumes were discarded for correction of the magnetic saturation effect, and the remaining volumes were first movement corrected and next slice-time corrected for temporal alignment. All voxels were spatially smoothed with a 6mm FWHM isotropic Gaussian kernel and after intensity normalization, a band pass filter was applied between 0.01 and 0.08 Hz (59), which was followed by the removal of linear and quadratic trends. We next regressed out the motion time courses, the average CSF signal, the average white matter signal and the average global signal. Finally, the functional data was spatially normalized to the MNI152 brain template.

**Further details on functional and structural data**

After data preprocessing, functional magnetic resonance imaging is based on the fact that brain activity variations are associated with changes in blood oxygenation. More specifically, the iron contained in hemoglobin (the protein transporting oxygen through the blood) is sensible to magnetic fields (63), allowing for tracking variations in blood oxygenation using fMRI (in-vivo and fully non-invasively). Hence, those time-series, based on the blood oxygenation level-dependent (BOLD) signal, provide an indirect measure of brain neuronal activity.

With regard to structural data, since the pioneer paper on diffusion tensor imaging in 1994 (64), diffusion tensor imaging (DTI) has been consolidated as the only method capable to non-invasively record in-vivo and the large-scale structural connectivity in human brain. The physics underlying DTI relies on the diffusion of water molecules, which occurs anisotropically through white matter tracts; such tracts constitute the physical skeleton providing brain structure. Importantly, DTI performance has been validated by replicating the SC extracted with other invasive methods; see e.g., a validation by post-mortem dissections of human brain (65), although the latter gave much higher resolution than DTI. Indeed, as conventional MRI equipment's can resolve diffusion times of about 50ms and water molecules diffuse within water with a coefficient of about $10^{-3}$ μm (66), this gives the broad estimation that DTI can capture free water molecules diffusion over distances ≈10 μm , which is about one order of magnitude bigger than the typical scale of a single cell. Thus, DTI can, by measuring the diffusion displacement of water molecules in the 3D space, obtain geometrical properties of the diffusion medium (i.e., axonal pathways). After preprocessing the raw data, DTI provides the tensor diffusion per each voxel of the 3D image.

**ROI extraction.** We applied the method of spatially constrained clustering to functional data averaged over the subjects (n=12) in order to extract the regions of interest (ROI), as explained in (60) and allowing for the generation of common ROIs. A spatial constraint is imposed to ensure that the resulting ROIs are spatially coherent and clustering was performed based on temporal correlations between voxel time series. To cluster at the group-level, a 2-level approach was applied in which the single subject data was first clustered and then all the subjects' data were combined to perform a second clustering. Finally, after the spatially constrained clustering, we applied a parcellation into 2514 ROIs in order partition the entire brain, including both cortical and sub-cortical regions (47% of the ROIs are cortical and 53% are sub-cortical, including the cerebellum).

**Calculation of structural and functional connectivity matrices.** For Structural connectivity (SC) matrices we computed the transformation from MNI152 brain template to individual fractional anisotropy maps. Using this transformation, the 2514 regions atlas was transformed to our diffusion image space. SC matrices were finally obtained by counting the number of fibers connecting each individual pair of ROIs. Functional connectivity (FC) matrices were calculated by obtaining the Pearson correlation coefficient between the rs-fMRI time series for each ROI pair.

**Common structure function modules (SFMs).** A hierarchical agglomerative clustering (HAC) was applied to extract brain modules on different scales, i.e.: SFMs. The first step is to select a set of features to describe each ROI. For rsFC, we employed the connectivity matrix of a given ROI as the feature vector to assess with all the other ROIs (2514 values for each ROI). Next, we applied the cosine distance to perform the clustering. For the SC, the feature vector was based on the distance, defined as one minus the fiber number normalized between 0 and 1. We also applied the cosine distance directly on the SC matrix and obtained similar results.

**Similarity (L) between SC and rsFC.** For each module of a given partition, the similarity between the corresponding sub-networks of SC and rsFC was calculated using the Sorensen index (61), but similar results were obtained using Jaccard similarity (figure S10). The Sorensen index accounts for the similarity between two binary datasets or clusters (i.e.: twice the number of common elements shared by the two modules divided by the total number of elements in the two modules). First, before making the rsFC and SC binary, we took the absolute rsFC value and we normalized the SC to values between zero and one. Next, we introduced two thresholds ranging from [0,1], $\alpha$ for rsFC and $\beta$ for SC, to binarize each rsFC or SC matrix according to whether a value was higher than the threshold, in which case the matrix element was equal to one (zero otherwise). For each module we then calculated the Sorensen index as a function of $\alpha$ and $\beta$. The values represented in figure S2a were obtained for the ($\alpha$, $\beta$) pair, maximizing the similarity for each module (typical values were about $\alpha = 0.45$ and $\beta$ very close to zero). Finally, we calculated the mean similarity across all modules.

Notice that the introduction of $\alpha$ and $\beta$ thresholds does indeed change the number of edges in the two graphs, but this preprocessing was necessary in order to have similar sparseness in the two matrices, i.e., whilst rsFC is practically a fully connected graph, SC is highly sparse. Such a preprocessing method guaranteed that the maximum similarity (in Sorensen's sense) occurred indeed when the two matrices had similar sparseness levels.

Finally, it is possible to address two different similarity values, one taking the HAC ordering of rsFC first and accordingly reordering the SC (blue line in figure S2a). Alternatively, the HAC ordering of SC can be taken, accordingly reordering the rsFC (red line in figure S2a).

**Modularity (Q) between rsFC and SC**. The Newman algorithm (62) was used to address modularity for a given brain partition. If M is the number of modules in the partition, modularity was calculated by $Q = \sum_{i=1}^{M}(e_{ii} - a_i^2)$, where $e_{ii}$ is the fraction of links connecting two ROIs that belong to the same module and $a_i$ is the fraction of links that connect a ROI from module i to other modules. Thus, partitions maximizing the within-module links and minimizing the between-modules links have high modularity values.

**Cross-modularity (X) between SC and rsFC.** We introduced the cross-modularity X to quantify both the topological similarity between rsFC and SC, and the individual modularities in rsFC and SC. Taking the HAC ordering in rsFC and reordering the SC, we defined cross-modularity between SC and rsFC as

$$X_{SF} = (Q_{FF} Q_{SF} \overline{L}_{SF})^{1/3}, \quad (1)$$

where $\overline{L}_{SF}$ is the mean similarity for all modules of the given partition, and $Q_{FF}$ and $Q_{SF}$ are the modularities of rsFC and SC, respectively (previously ordered with the HAC ordering of rsFC). Similarly, taking the HAC ordering of SC and reordering the rsFC, the cross-modularity is:

$$X_{FS} = (Q_{SS} Q_{FS} \overline{L}_{FS})^{1/3}. \quad (2)$$

**Percentage Overlap.** To identify the SFMs extracted with other brain regions (cf. figures 6, S6 and S7), we calculated the overlap between the SFMs and the AAL brain areas (44), the resting state networks (45) and the Brodmann areas (included in the MRIcro software http://www.mricro.com). Overlapping was addressed using the Sorensen index as defined previously for similarity. Statistical significance was addressed by generating 100 random permutations of a given brain partition and the p-values were calculated by the cumulative distribution of the Gaussian distribution. Only similarity values with a p-value less than 0.05 were considered.

**Acknowledgments**

Work supported by Ikerbasque: The Basque Foundation for Science, Euskampus at UPV/EHU, Gobierno Vasco (Saiotek SAIO13-PE13BF001) and Junta de Andalucía (P09-FQM-4682) to JMC; Ikerbasque Visiting Professor to SS; Junta de Andalucía (P09-FQM-4682) and Spanish Ministry of Economy and Competitiveness (FIS2013-43201-P) to MAM; the European Union's Seventh Framework Programme (ICT-FET FP7/2007-2013, FET Young Explorers scheme) under grant agreement n. 284772 BRAIN BOW (www.brainbowproject.eu) and by the Joint Italy—Israel Laboratory on Neuroscience to PB.

For results validation (figure S8), data were provided by the Human Connectome Project, WU-Minn Consortium (Principal Investigators: David Van Essen and Kamil Ugurbil; 1U54MH091657) funded by the 16 NIH Institutes and Centers that support the NIH Blueprint for Neuroscience Research; and by the McDonnell Center for Systems Neuroscience at Washington University.

## Additional Information

**Author contributions.** IE and BM performed MRI acquisitions; ID preprocessed and postprocessed neuroimaging data; ID and PB performed data analysis; ID, PB, MAM, SS and JMC wrote the manuscript; ID, PB, MAM, SS and JMC prepared the figures; SS and JMC led the research; all the authors reviewed the manuscript.

**Competing Financial Interest.** The authors declare no competing financial interests in relation to this work.

**Figure Legends**

**Fig. 1: Strong similarities between structural and functional brain networks emerge out of a hierarchical modular organization. A:** The top row of images are the rsFC (left) and SC (right) matrices averaged over 12 healthy patients with 2514x2514 ROIs (47% of them are cortical ROIs, 53% sub-cortical, including the cerebellum). M=20 brain modules were identified by applying hierarchical agglomerative clustering to the rsFC matrix, plotting the matrix elements of SC (binarized for clarity of visualization) in the same order as those in the rsFC. Amplification of the diagonal of the rsFC and SC are presented in different colored rectangles (red, ochre, blue and green). The SC matrix shows strong similarities to the modular organization of rsFC, even though the SC is sparser than the rsFC. **B:** The M=20 modules are colored on top of the SC. **C:** The dendrogram applied to the rsFCs identifying the M=20 modules. **D:** A 3D brain representation of all 20 modules. **E:** As an example, we plotted the module 2 to show that our brain partition chooses modules satisfying that the different functionally correlated brain areas (colored in maroon) belonging to the same module are also structurally connected. A 3D movie of module 2 is given in movie S22.

**Fig. 2: Cross-modularity index (X) between rsFC and SC.** Cross-modularity has been calculated for brain partitions of different sizes, varying from 1 (the entire brain) to 100 modules. The cross-modularity, a novel index introduced here for the first time, increases when either the topological moduli-similarity between rsFC and SC increases or if the individual modularity in rsFC or SC does it (see SOM). A stronger cross-modularity between rsFC and SC was achieved by applying the HAC to the rsFC (blue curve $X_{SF}$) rather than to the SC (red curve $X_{FS}$). The arrow at M=20 indicates that at that point, both the blue and red curves are well represented in the hierarchical agglomerative clustering, with an optimal cross-modularity index.

**Fig. 3: Visualization of the common structure-function modules (SFMs)**. A brain atlas of 20 SFMs that maximizes the cross-modularity index (here only represented modules from 1 to 10, and similarly, figure 4 showing modules from 11 to 20). These networks were obtained by identifying the 20 modules in the rsFC matrix. Note that some of the modules are composed of spatially separated brain regions (e.g., # 2, 9 in figure 3 and 12, 14, 16, 17 and 18 in figure 4). 3D movies of the 20 modules are available in supplementary material (movies S1-S20). Movie S21 corresponds to a 3D superposition of all the 20 modules. In Table S1, we also provide a detailed anatomical description of the 20 modules.

**Fig. 4: Visualization of the common structure-function modules (SFMs)**. See caption of figure 3.

**Fig. 5: Positive (red), negative (blue) and near-zero (green) correlations in rsFC.** The figure shows that SFMs are internally correlated (red) and that many of them tended to be anti-correlated with others (blue). Black numbers are indicating different SFM number for reference purposes.

**Fig 6: Percentage overlap between SFMs and previously described brain parcellations. A:** The anatomical brain partition described in the AAL atlas. **B:** The resting state networks (RSNs). **A,B:** Percentage overlap between the M=20 SFMs and the specific parcel.

**Figure 1:**

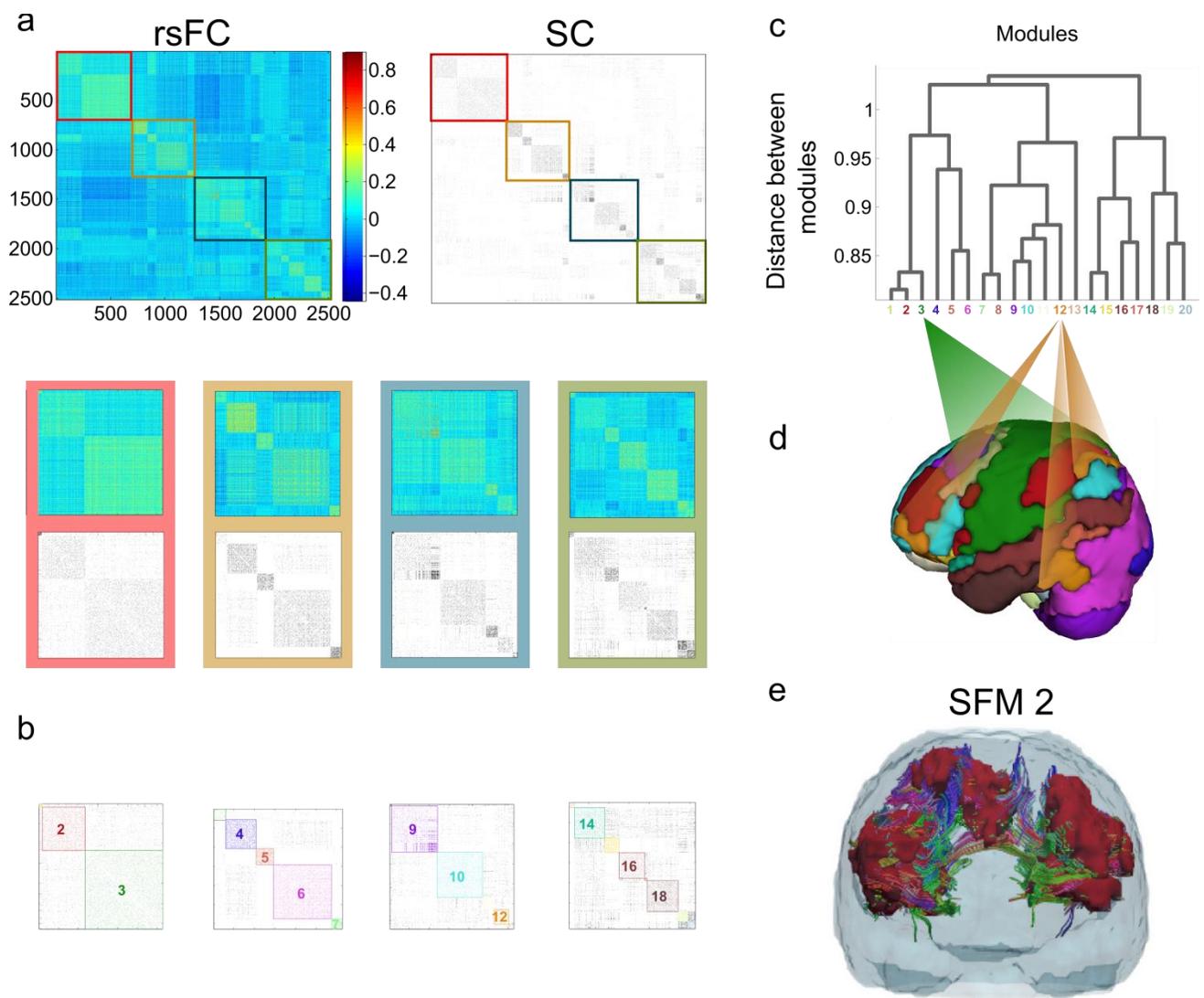

**Figure 2:**

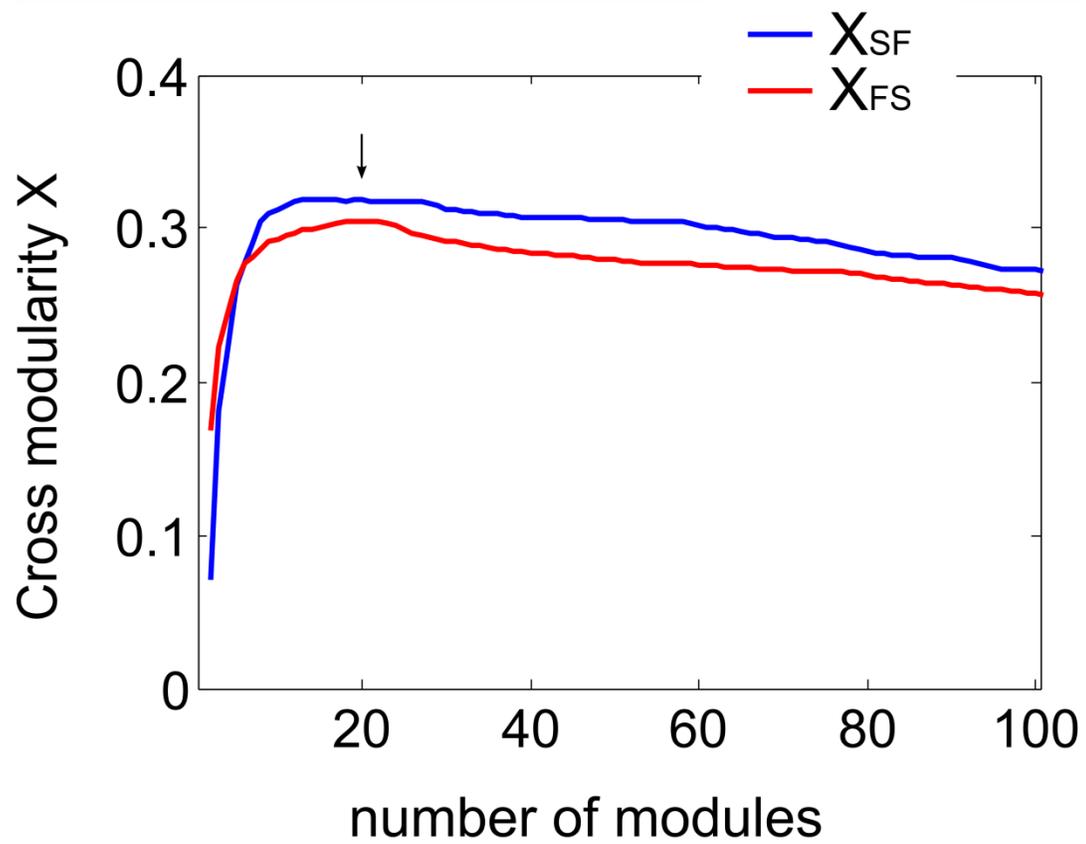

**Figure 3:**

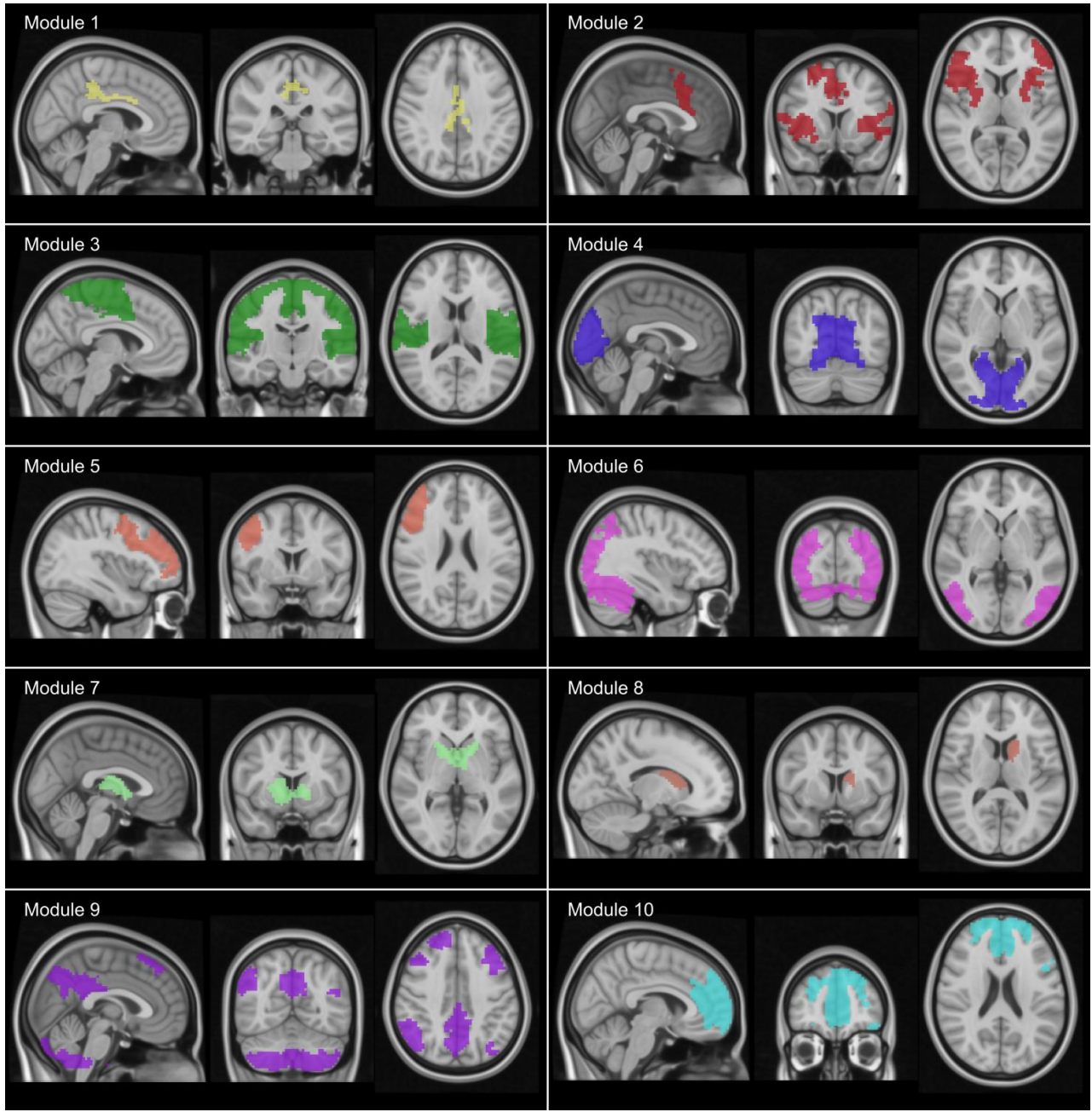

**Figure 4:**

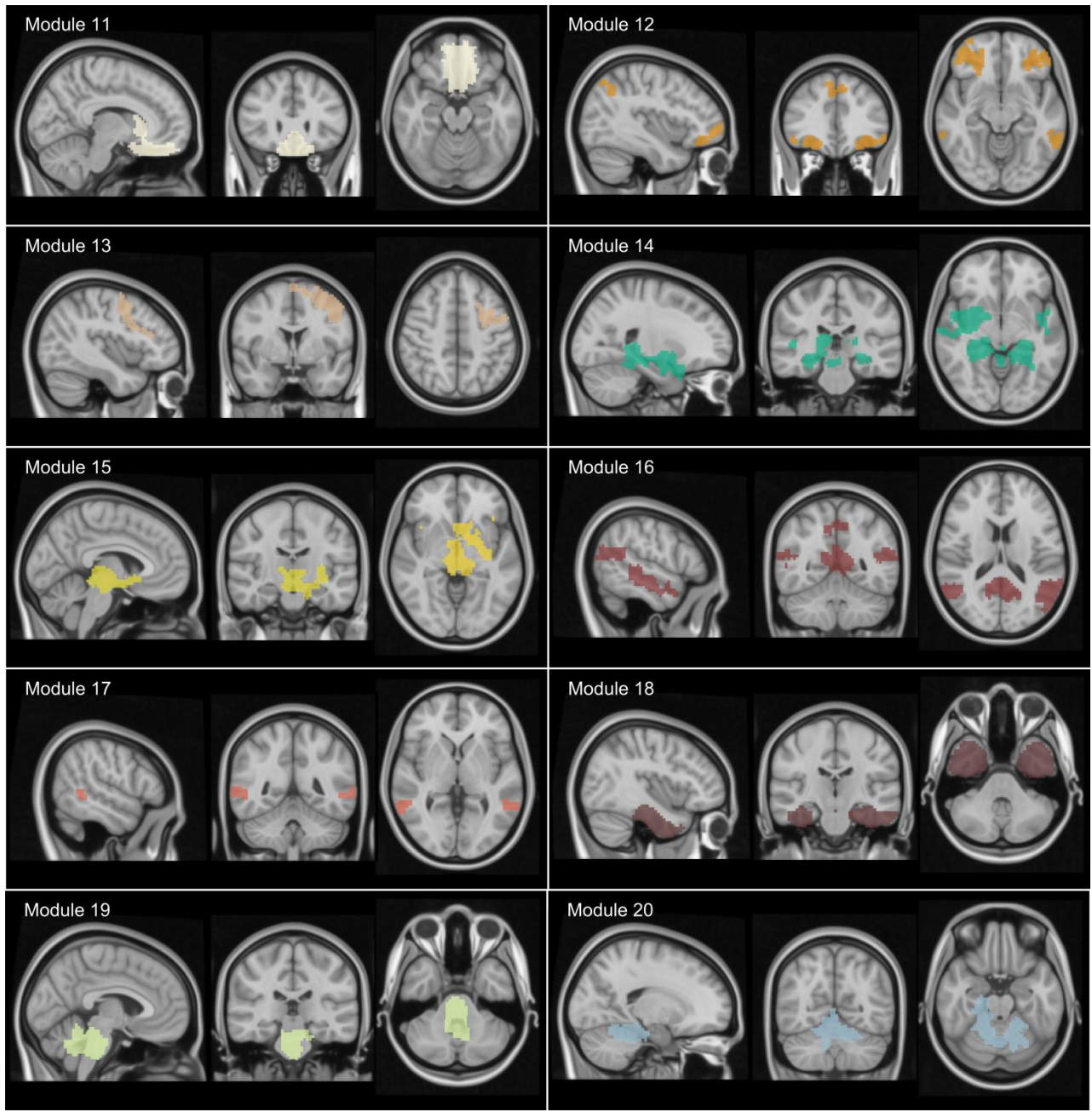

**Figure 5:**

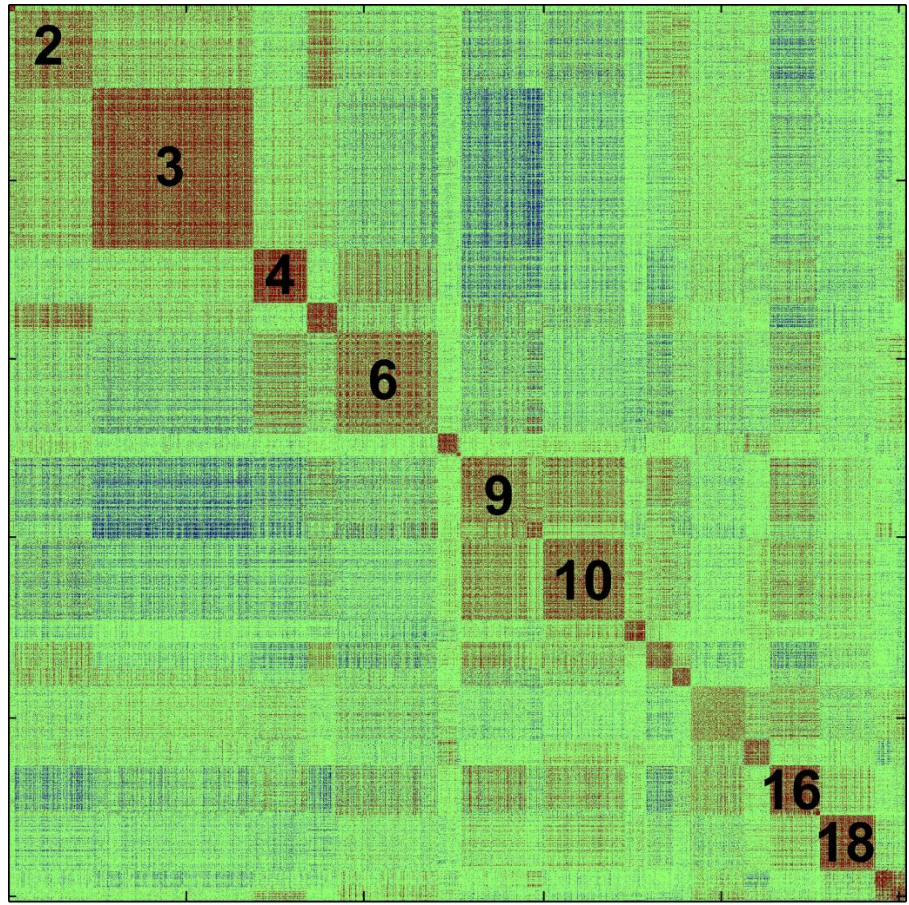

**Figure 6:**

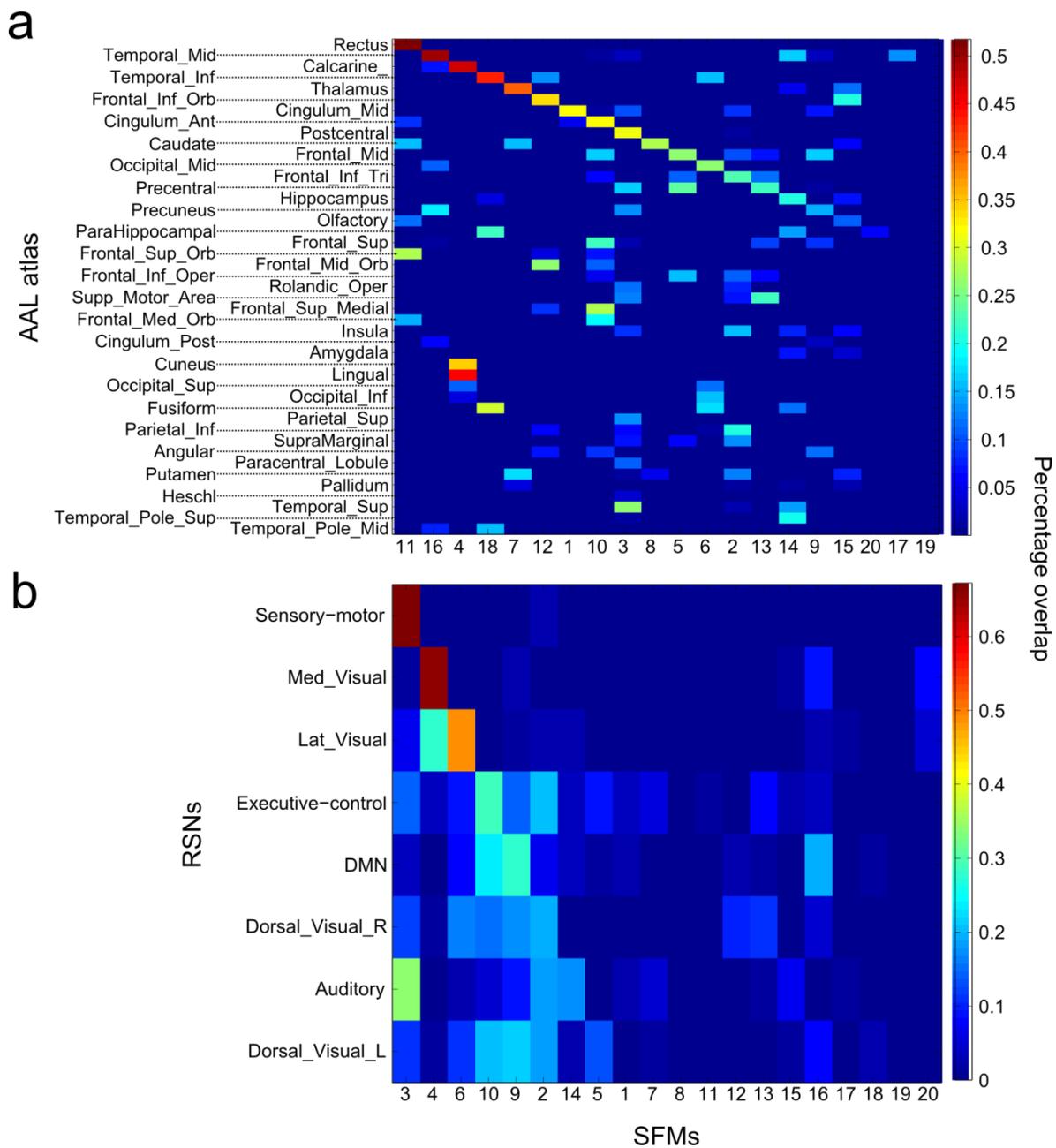

# Supplementary Information:
# A novel brain partition highlights the modular skeleton shared by structure and function

Ibai Diez[1,*], Paolo Bonifazi[2,*], Iñaki Escudero[1,3], Beatriz Mateos[1,3], Miguel A. Muñoz[4], Sebastiano Stramaglia[1,5,6,**,§] and Jesus M Cortes[1,6,7,§]

1. Biocruces Health Research Institute, Cruces University Hospital, Barakaldo, Spain.

2. School of Physics and Astronomy, George S. Wise Faculty of Life Sciences, Sagol School of Neuroscience, Tel Aviv University, Tel Aviv, Israel.

3. Radiology Service, Cruces University Hospital, Barakaldo, Spain.

4. Departamento de Electromagnetismo y Física de la Materia and Instituto Carlos I de Física Teórica y Computacional, Universidad de Granada, Spain.

5. Dipartimento di Fisica, Universita degli Studi di Bari and INFN, Bari, Italy.

6. Ikerbasque: The Basque Foundation for Science, Bilbao, Spain.

7. Department of Cell Biology and Histology. University of the Basque Country. Leioa, Spain

* These authors contributed equally to this work.

** New address: BCAM - Basque Center for Applied Mathematics, Bilbao, Spain

§ To whom correspondence should be addressed.

Email: Sebastiano.Stramaglia@ba.infn.it and jesus.cortesdiaz@osakidetza.net

**Table S1: Anatomical description of the 20 modules defined after hierarchical agglomerative clustering (HAC) of rsFC.** In the first column, we also provide the module volume and links to the 3D movies are given in the third column.

| Module number (volume size) | Anatomical description |
|---|---|
| module 1 (7.26 cm$^3$) | **Posterior cingulate:** posterior area of the cingulate gyrus or callosal convolution. Located above the corpus callosum, it goes from the frontal lobe back to the temporal uncus and up to the splenium. It belongs to the Default Mode Network. |
| module 2 (104.36 cm$^3$) | **Putamen:** a round structure located at the base of the telencephalon. It is also one of the basal ganglia structures.<br>**Anterior cingulate:** anterior frontal region of the cingulate gyrus, initiated above the rostrum of the corpus callosum.<br>**Rostral pars of the middle frontal gyrus:** anterior inferior end of the middle frontal gyrus.<br>**Superior parietal gyrus:** parietal gyrus located posterior to the postcentral gyrus.<br>**Supramarginal gyrus**: region in the parietal lobe encircling the posterior extreme of the Sylvian fissure.<br>**Insula:** triangular area of cerebral cortex forming the medial wall of the Sylvian fissure.<br>**Inferior parietal gyrus:** parietal gyrus located behind the postcentral gyrus and below the superior parietal gyrus.<br>**Precentral gyrus:** frontal gyrus that defines the anterior boundary of the fissure of Rolando with a mainly motor function.<br>**Superior frontal gyrus:** antero-superior parasagittal frontal gyrus, located anterior to the precentral gyrus. |
| module 3 (221.18 cm$^3$) | **Paracentral lobule:** medial gyrus that connects the pre- and postcentral gyrus.<br>**Precentral gyrus** (cf. Module 2)<br>**Postcentral gyrus:** Parietal gyrus located between the fissure of Rolando and the postcentral sulcus, which has a mainly sensory function.<br>**Precuneus:** square brain lobule located before the parietal-occipital sulcus and behind the paracentral lobule at the medial surface of the brain hemisphere.<br>**Superior frontal gyrus** (cf. module 2).<br>**Superior parietal gyrus** (cf. module 2)<br>**Superior temporal gyrus:** temporal gyrus at the lateral surface of the |
| | temporal lobe. It is located below the Sylvian fissure and above the superior temporal sulcus. It belongs to the temporal neocortex.<br>**Supramarginal gyrus** (cf. module 2).<br>**Insula** (cf. module 2) |
| module 4<br><br>(91.48 cm$^3$) | **Cuneus:** occipital gyrus between the parieto-occipital sulcus and the calcarine sulcus at the medial surface of the occipital lobe.<br>**Lateral occipital sulcus:** external lateral surface of the occipital lobe close to the occipital lobe, dividing the external occipital gyrus.<br>**Lingual gyrus:** occipital extension of the parahippocampal gyrus at the medial surface of the occipital lobe.<br>**Pericalcarine cortex:** occipital area encircling the calcarine sulcus with a function associated to visual perception.<br>**Precuneus** (cf. module 3) |
| module 5<br><br>(37.02 cm$^3$) | **Medial frontal gyrus:** frontal gyrus at the lateral surface below the superior frontal gyrus.<br><br>**Precentral gyrus** (cf. module 2)<br><br>**Rostral pars of the middle frontal gyrus** (cf. module 2) |
| module 6<br><br>(159.33 cm$^3$) | **Cerebellum:** posterior part of the rombencephalon made up of the two hemispheres and the central vermis. It is located below the occipital lobe.<br><br>**Fusiform gyrus:** temporal gyrus in the inferior surface between the inferior temporal gyrus and the parahippocampal gyrus. It has two areas, the medial occipito-temporal gyrus and the lateral occipito-temporal gyrus.<br><br>**Inferior temporal gyrus:** inferior gyrus located in the lateral surface of the temporal lobe, below the inferior temporal sulcus.<br><br>**Lateral occipital sulcus** (cf. module 4)<br><br>**Superior parietal gyrus** (cf. module 2) |
| module 7<br><br>(22.30 cm$^3$) | **Thalamus:** middle symmetrical structure of the diencephalon with multiple afferent and efferent connections, situated around the third ventricle.<br><br>**Caudate nucleus** (symmetrical structure): one of the basic structures of the basal ganglia belonging to the corpus striatum. It is located at the lateral surface of the lateral ventricles surrounding the thalamus. |
|  | **Putamen** (cf. module 2) |
|  | **Pallidum:** symmetrical structure within the basal ganglia. Medial diencephalic region of the lenticular nucleus. |
|  | **Accumbens nucleus:** symmetrical structure located in the ventral region of the corpus striatum, therefore belonging to the basal ganglia. |
| module 8 (3.29 cm$^3$) | **Caudate nucleus** (cf. module 7) |
|  | **Putamen** (cf. module 2) |
| module 9 (163.67 cm$^3$) | **Cerebellum** (cf. module 6) |
|  | **Caudal middle frontal:** frontal gyrus on the lateral surface, located below and lateral to the superior frontal gyrus. This region refers to its most caudal part. |
|  | **Cingulate isthmus:** intersection narrowing between the cingulate and the hippocampal gyrus. It is located behind and below the splenium of corpus callosum. |
|  | **Posterior cingulate** (cf. module 1) |
|  | **Precuneus** (cf. module 3) |
|  | **Inferior parietal gyrus** (cf. module 2) |
|  | **Rostral pars of the middle frontal gyrus** (cf. module 2) |
|  | **Superior frontal gyrus** (cf. module 2) |
| module 10 (103.55 cm$^3$) | **Anterior cingulate** (cf. module 2) |
|  | **Inferior parietal gyrus** (cf. module 2) |
|  | **Orbital gyrus:** frontobasal gyrus lateral located to the straight gyrus. |
|  | **Pars opercularis:** opercular part of the inferior frontal gyrus. |
|  | **Pars orbitalis:** orbital part of the inferior frontal gyrus. |

|  | **Pars triangularis:** inferior part of the inferior frontal gyrus. |
|  | **Anterior cingulate** (cf. module 2) |
|  | **Rostral pars of middle frontal gyrus** (cf. module 2) |
|  | **Superior frontal gyrus** (cf. module 2) |
| module 11<br>(31.08 cm$^3$) | **Caudate nucleus** (cf. module 7) |
|  | **Accumbens nucleus** (cf. module 7) |
|  | **Lateral frontal orbital gyrus:** external orbital gyrus, located frontobasal and lateral to the medial orbitofrontal gyrus. |
|  | **Orbital gyrus** (cf. module 10)<br>**Anterior cingulate** (cf. module 10) |
| module 12<br>(33.24 cm$^3$) | **Inferior parietal gyrus** (cf. module 2) |
|  | **Inferior temporal gyrus** (cf. module 6) |
|  | **Lateral frontal orbital gyrus** (cf. Module 11) |
|  | **Pars orbitalis** (cf. module 10) |
|  | **Pars triangularis** (cf. module 10) |
|  | **Rostral pars of the middle frontal gyrus** (cf. module 2) |
|  | **Superior frontal gyrus** (cf. module 2) |
|  | **Caudate nucleus and anterior cingulate** (cf. module 7 and module 2) |
| module 13<br>(24.46 cm$^3$) | **Middle frontal gyrus:** caudal part of the middle frontal gyrus.<br>**Pars opercularis** (cf. module 10)<br>**Precentral gyrus** (cf. module 2)<br>**Superior frontal gyrus** (cf. module 2) |
|  |  |
| module 14<br><br>(92.75 cm$^3$) | **Thalamus** (cf. module 7)<br><br>**Hippocampus:** symmetrical grey matter structure, located in the mesial-temporal region, at the base of the temporal horn.<br><br>**Amygdala:** grey nuclei located in the temporal uncus, above the temporal ventricular horn. It belongs to the rhinencephalon.<br><br>**Putamen** (cf. modulo 2)<br><br>**Ventral diencephalon:** multiple structures containing the hypothalamus, mammillary tubercle, subthalamic nucleus, substantia nigra, red nucleus, geniculate body, optic tract and cerebral peduncles.<br><br>**Banks of the superior temporal sulcus:** Temporal lobe structure between the superior temporal gyrus and the middle temporal gyrus.<br><br>**Parahippocampal gyrus:** convolution located below the hippocampal sulcus in the temporal mesial region.<br><br>**Superior temporal gyrus** (cf. module 3)<br><br>**Insula** (cf. module 2)<br><br>**Middle temporal gyrus:** gyrus located on the lateral surface of the temporal lobe between the inferior and superior temporal sulcus.<br><br>**Temporal pole:** anterior end of the temporal lobe. |
| module 15<br><br>(42.96 cm$^3$) | **Thalamus** (cf. module 7)<br><br>**Putamen** (cf. module 2)<br><br>**Pallidum** (cf. module 7)<br><br>**Brainstem:** it consists of three parts, the myelencephalon, pons (metencephalon) and midbrain (mesencephalon). It is the main communication route between the brain, spinal cord and peripheral nerves.<br><br>**Hippocampus** (cf. module 14)<br><br>**Amygdala** (cf. module 14)<br><br>**Accumbens nucleus** (cf. module 7) |

|  | **Ventral diencephalon** (cf. module 14) |
|  | **Orbital gyrus (cf. module 10)** |
|  | **Insula** (cf. module 2) |
| module 16<br>(65.58 cm$^3$) | **Cerebellum** (cf. module 6) |
|  | **Banks of the superior temporal sulcus** (cf. module 14) |
|  | **Inferior parietal gyrus** (cf. module 2) |
|  | **Cingulate isthmus** (cf. module 9) |
|  | **Middle temporal gyrus** (cf. module 14) |
|  | **Precuneus** (cf. module 3) |
|  | **Superior temporal gyrus** (cf. module 3) |
| module 17<br>(5.29 cm$^3$) | **Banks of the superior temporal sulcus** (cf. module 14) |
|  | **Middle temporal gyrus** (cf. module 14) |
| module 18<br>(74.39 cm$^3$) | **Hippocampus** (cf. module 14) |
|  | **Amygdala** (cf. module 14) |
|  | **Entorhinal cortex**: area in the medial-temporal lobe located between the hippocampus and temporal neocortex. |
|  | **Fusiform gyrus** (cf. module 6) |
|  | **Inferior temporal gyrus** (cf. module 6) |
|  | **Middle temporal gyrus** (cf. module 14) |
|  | **Parahippocampal gyrus** (cf. module 14) |
|  | **Temporal pole** (cf. module 14) |

| module 19 (28.54 cm$^3$) | **Cerebellum** (cf. module 6) <br> **Brainstem** (cf. module 15) |
|---|---|
| module 20 (34.91 cm$^3$) | **Cerebellum** (cf. module 6) <br> **Parahippocampal gyrus** (cf. module 14) |

**Figure S1:**

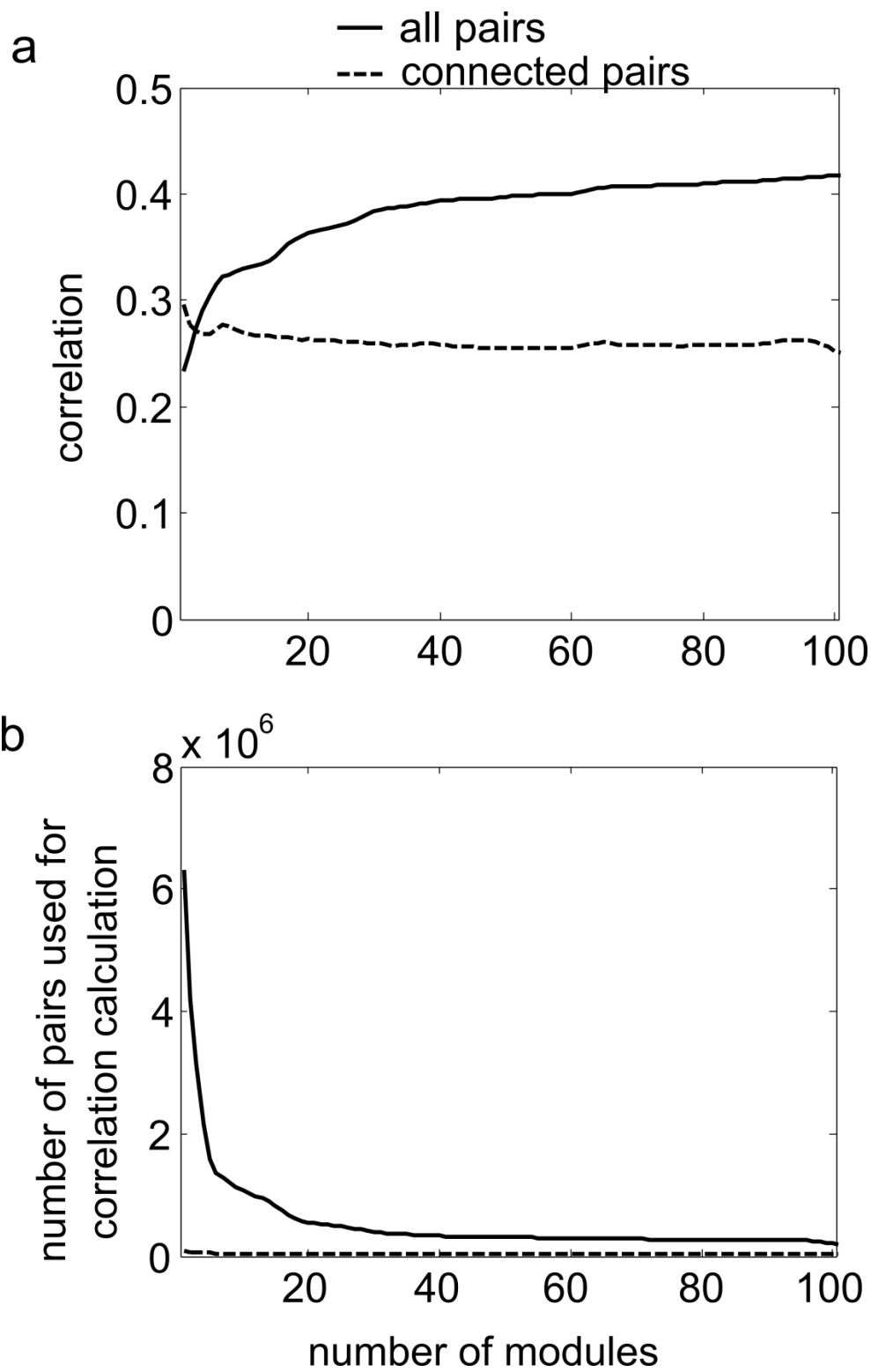

**Figure S2:**

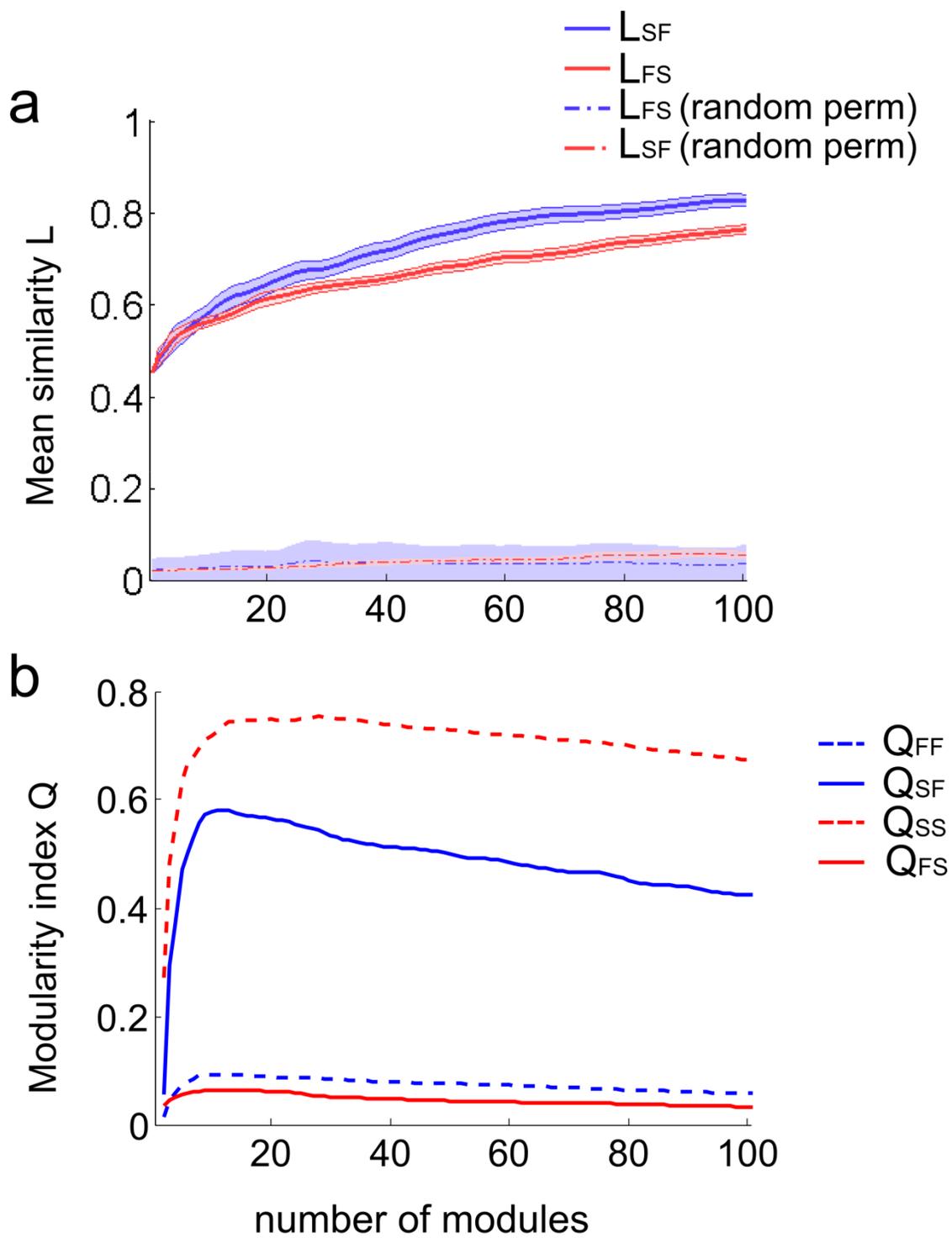

**Figure S3:**

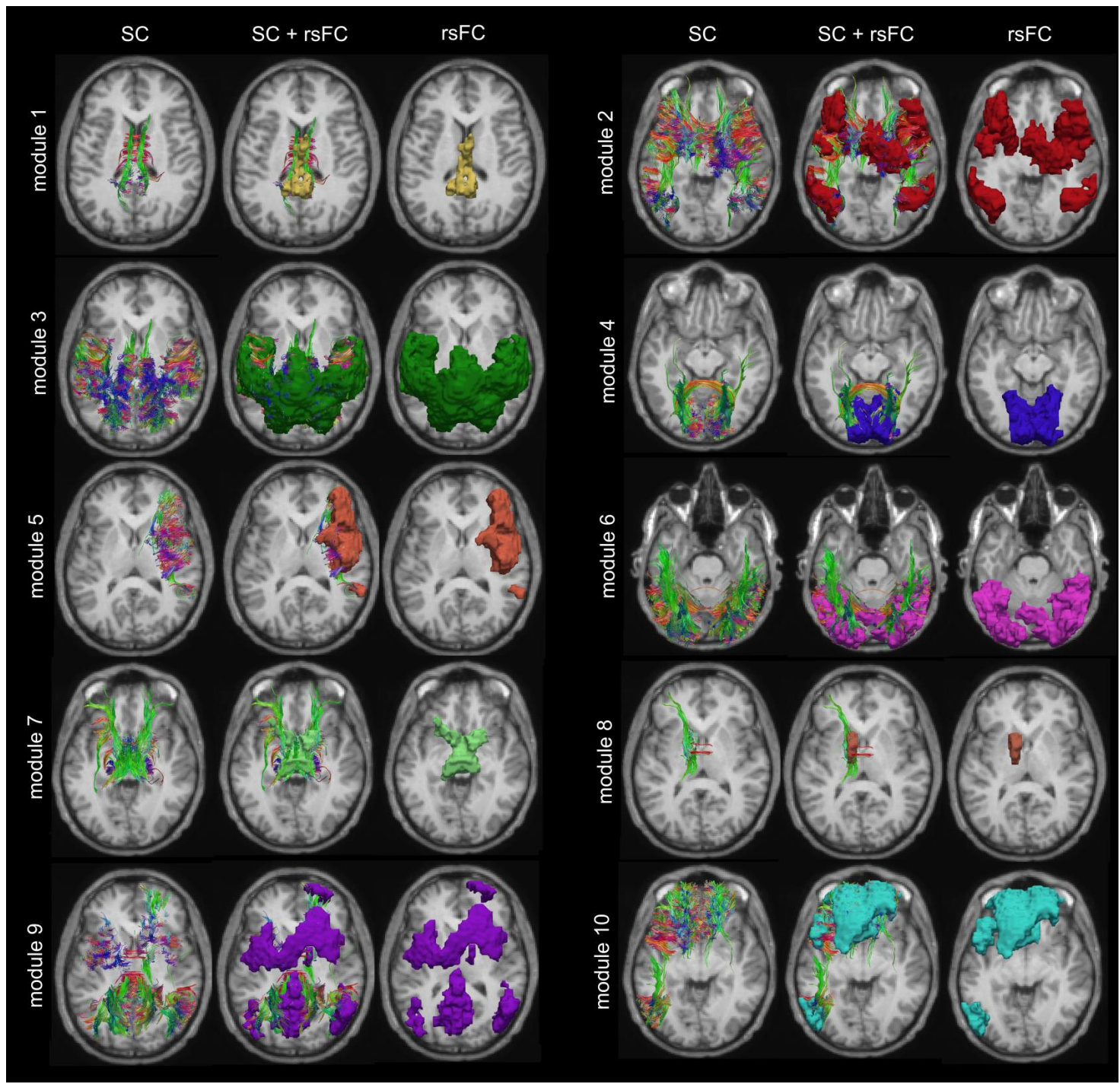

**Figure S4:**

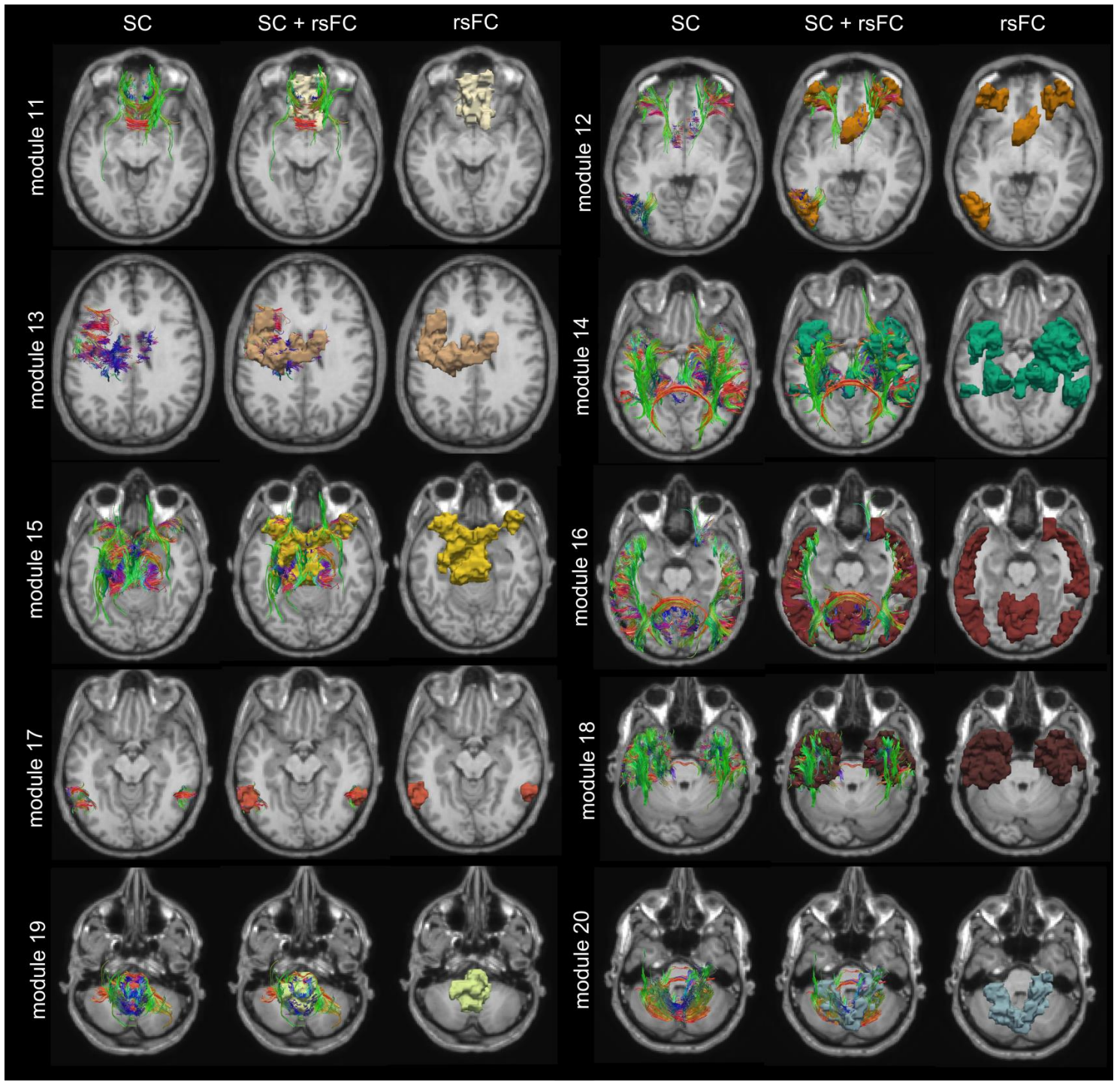

**Figure S5:**

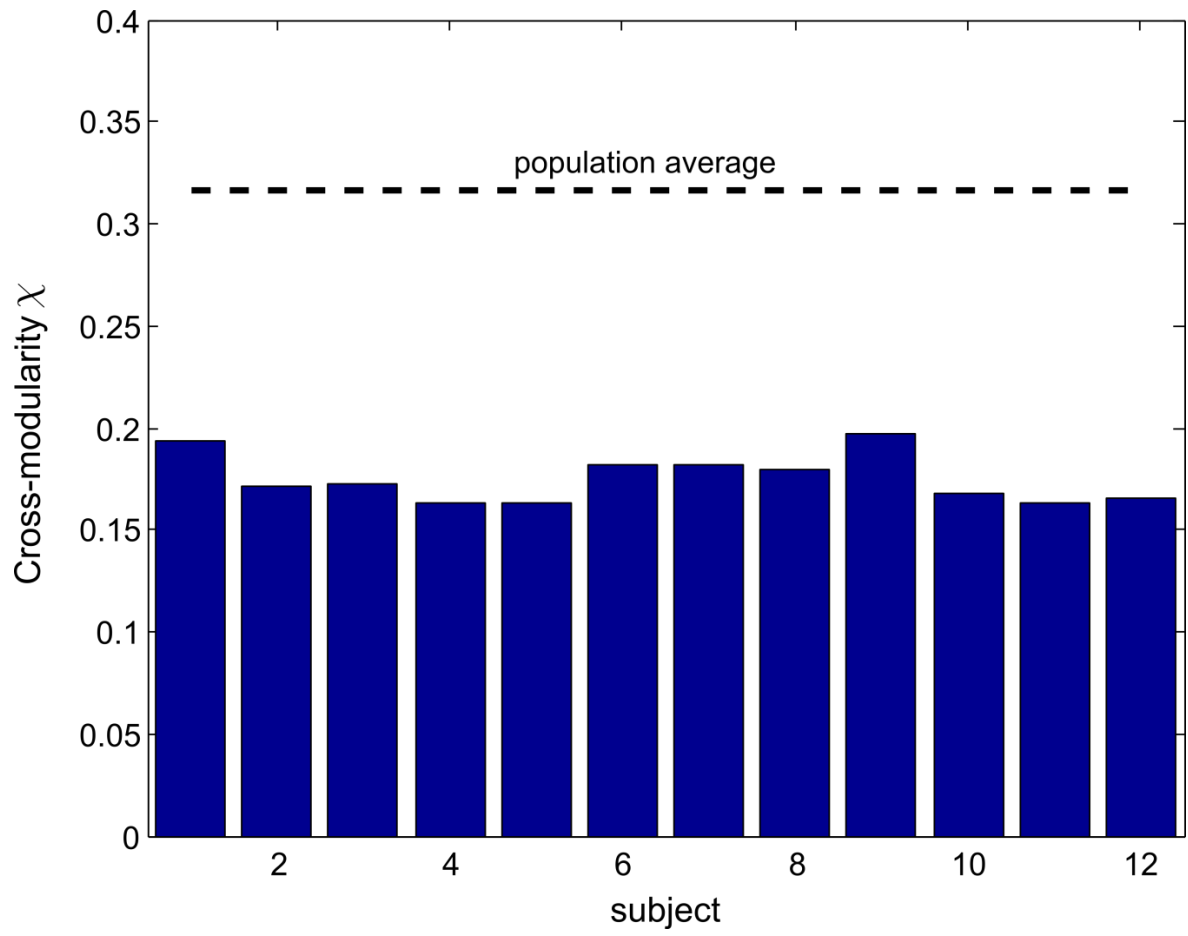

**Figure S6:**

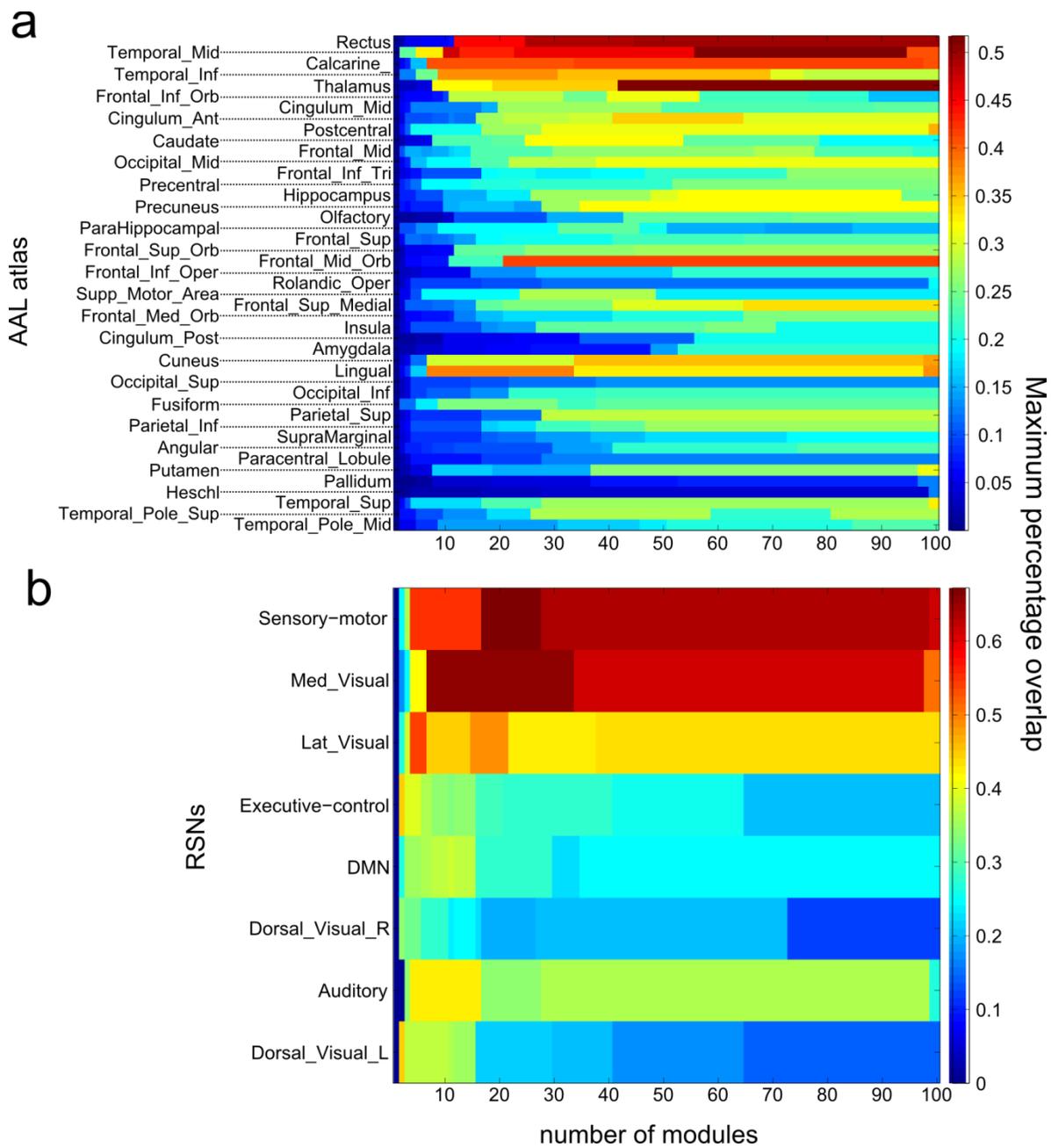

**Figure S7:**

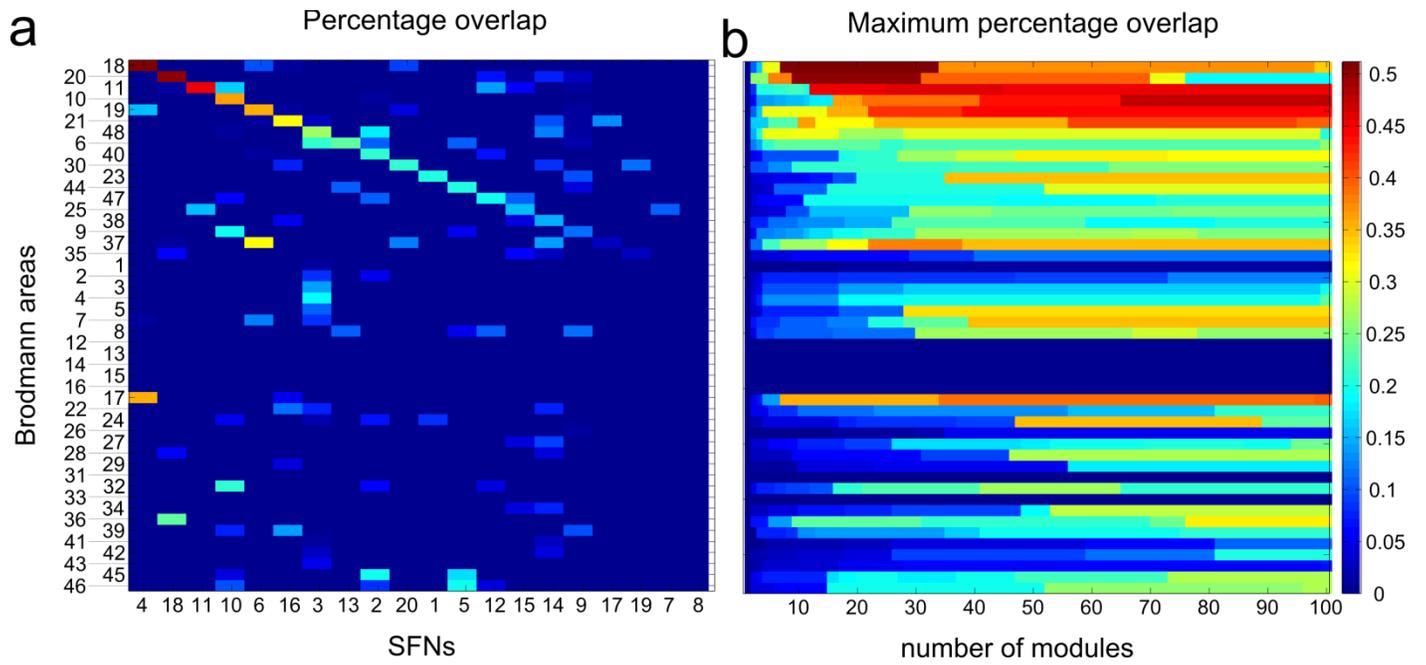

**Figure S8:**

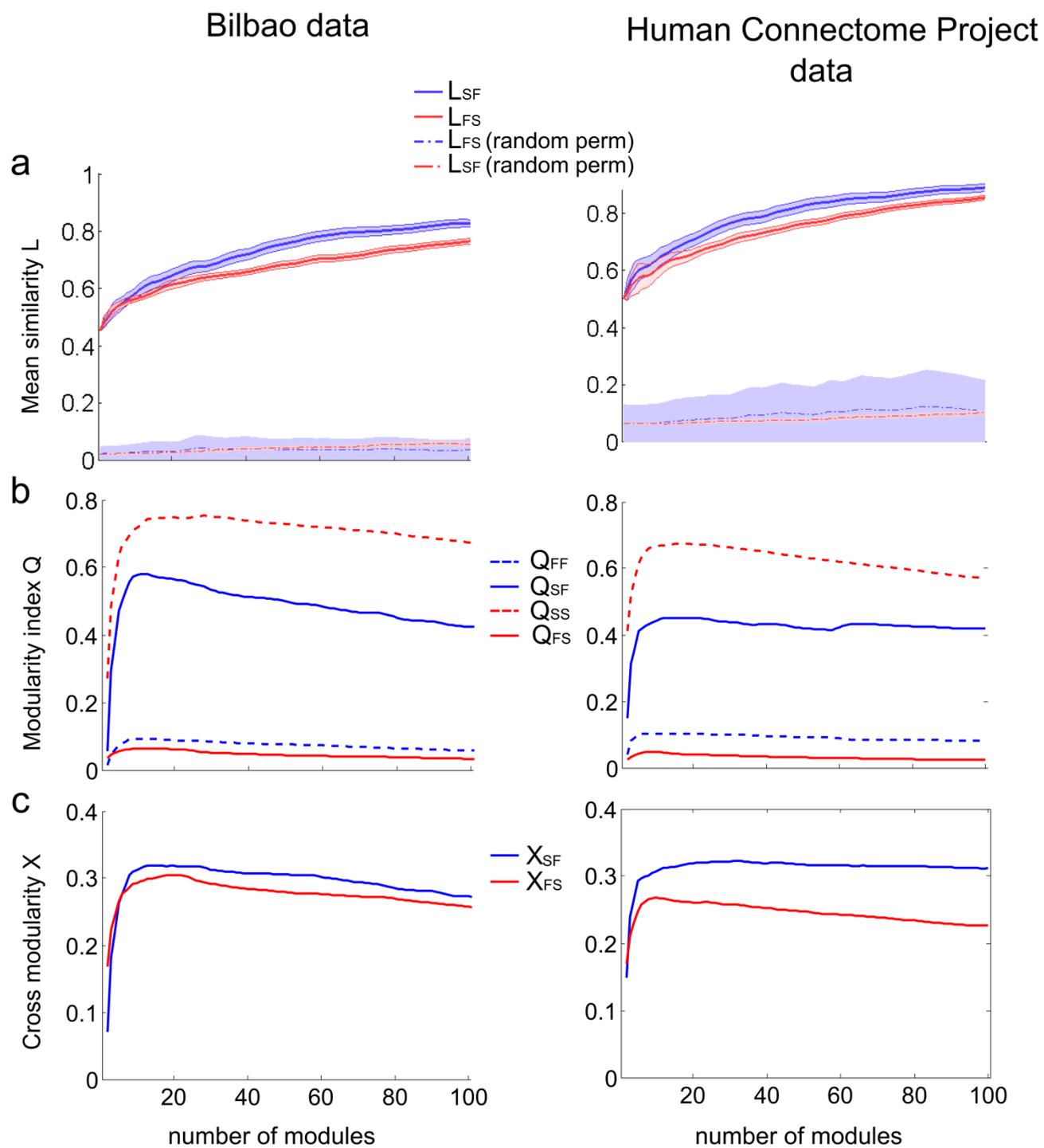

**Figure S9:**

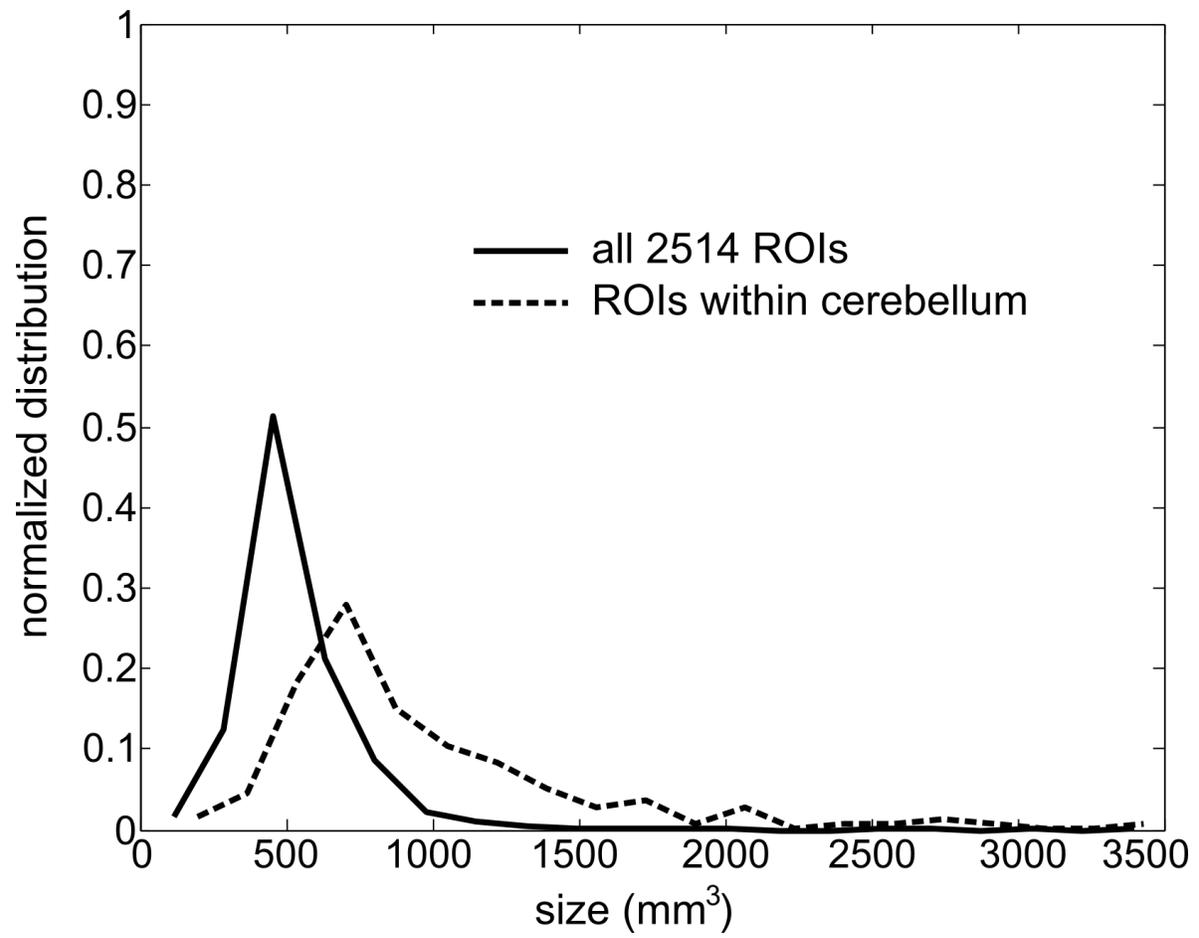

**Figure S10:**

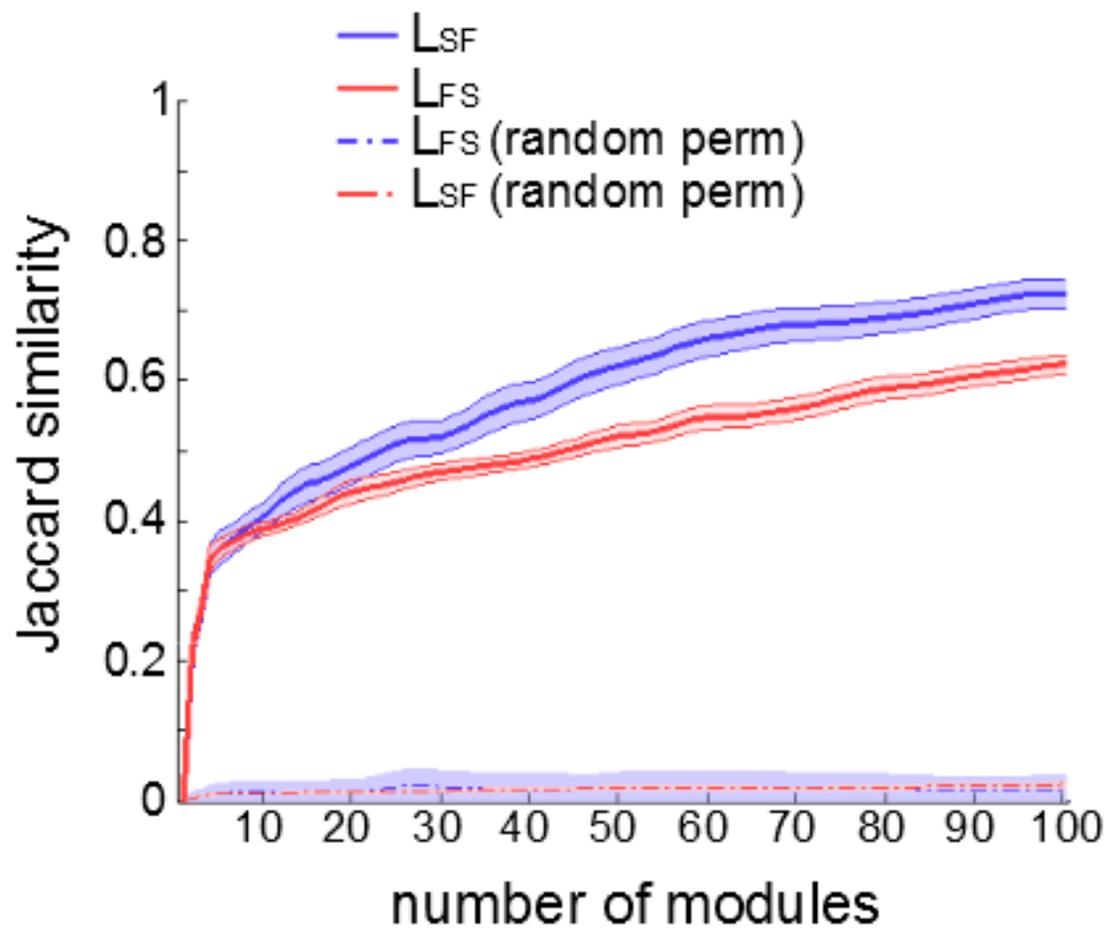

**Supplementary Figure Legends**

**Fig. S1: Pearson correlation between rsFC and SC. A:** Similar to the pioneering works on matching rsFC and SC by O. Sporns group, we calculated the Pearson correlation value between rsFC and SC (first calculated using all matrix-elements in rsFC and SC). The situation of a single module in the brain partition recovers that indicated by previous authors. Similarly, for a single module in the brain partition, the correlation calculated on connected pairs ("pairs with fiber number >0", dashed line) gave a higher value in comparison to all the pairs (solid line). As the number of the modules in the brain partition increases, we represent the correlation between rsFC and SC, yet only on pairs belonging to same moduli in rsFC and SC (after HAC on rsFC, we accordingly reordered the SC). **B:** the number of pairs taken into account for the calculation of the correlation in function of the number of modules.

**Fig. S2: Similarity (L) and Modularity (Q) between rsFC and SC. A:** The mean similarity (± SE) between functional and structural moduli was calculated for brain partitions of different sizes varying from 1 (the entire brain) to 100 modules (similarly in panel B). The $L_{SF}$ was obtained by applying hierarchical agglomerative clustering to rsFC and reordering the SC accordingly. The $L_{FS}$ represents the specular case and the dashed lines correspond to the mean similarity in the case of random permutations in rsFC (blue) and SC (red). **B:** $Q_{FF}$ and $Q_{SF}$ refer to the respective modularity of rsFC and SC on the brain modules as achieved by applying HAC to rsFC. Similarly, $Q_{SS}$ and $Q_{FS}$ are the respective modularity of SC and rsFC on the brain modules achieved by applying HAC to SC.

**Fig. S3: SFMs illustrating common modularity structure between SC and rsFC.** From module 1 to 10, we are plotting modules obtained only with SC (left column), with both SC and rsFC (middle column) and only with rsFC (right column).

**Fig. S4: SFMs illustrating common modularity structure between SC and rsFC.** Similar to figure S3 but for modules from 11 to 20.

**Fig. S5: Cross-modularity between rsFC and SC across different subjects.** The dashed line represents the cross-modularity between the average population (n=12) rsFC and SC.

**Fig. S6: Maximum percentage overlap between SFMs and previously described brain parcellations. A:** Maximum percentage overlap between each region in the AAL and all the different SFMs after HAC, varying the number of modules. **B:** similar to A, but for RSNs.

**Fig. S7: Percentage overlap between SFMs and the Brodmann areas.** Similar to figures 6 and S6 but for Brodmann areas.

**Fig. S8: Validation of our results using data from the Human Connectome Project**. Maximization of cross-modularity, on both our data (Bilbao) and WU-Minn Human Connectome Project lead to almost an identical brain partition.

**Fig. S9: Distribution of ROIs' size.** The solid line corresponds to the size (measured in mm3) distribution for all the 2514 ROIS. The dashed line does the same but for the ROIs located within the cerebellum (a number of 211 to the total 2514 ROIs).

**Fig. S10: Jaccard similarity between SC and rsFC as a function of the number of modules.** The results are very similar to those obtained using Sorensen's similarity (fig. S2).